\documentclass{article}

\usepackage[english]{babel}

\usepackage[utf8]{inputenc}
\usepackage[normalem]{ulem}
\usepackage{amsmath, amssymb, amsfonts, amsbsy, amscd, mathtools, amsthm}
\usepackage{latexsym}
\usepackage{dsfont}
\usepackage{hyperref}
    \hypersetup{
        colorlinks=true,
        linkcolor=blue,
        citecolor=red,
    }

\usepackage[sort,nocompress]{cite}

\usepackage[margin=2.5cm]{geometry}
\usepackage{setspace}                       % adjust line spacing
      
\usepackage{xcolor}
    \definecolor{applegreen}{rgb}{0.55, 0.71, 0.0}
    \definecolor{anothergreen}{rgb}{0.55, 0.60, 0.0}
    \definecolor{myblue}{rgb}{0.0352,0.4981,0.6509}
    \definecolor{redbrick}{rgb}{0.82, 0.1, 0.26}
    
\usepackage{longtable, array, booktabs, tabularx}   % table packages
\usepackage{subcaption}    
\usepackage{graphicx}                       % images

\usepackage{tikz,pgfplots}                  % TikZ drawings
    \pgfplotsset{compat=1.18}

\newcommand{\commentout}[1]{}
\newcommand {\bv} { {\mathbf{v}} }

\newcommand {\bx} { {\mathbf{x}} }

\newcommand{\beq}{\begin{equation}}
\newcommand{\eeq}{\end{equation}}
\newcommand{\bea} {\begin{array}{rl}}
\newcommand{\eea} {\end{array}}
\newcommand{\bepa}{\left\{ \begin{array}{l}}
\newcommand{\eepa} {\end{array}\right.}
\newtheorem{theorem}{Theorem}%[section]

\newtheorem{remark}[theorem]{Remark}

\title{The influence of uninformed individuals in opinion-swarming models for self-propelled particle system}
\title{Decision making in heterogeneous self-propelled particle systems}

\author{Gissell Estrada-Rodriguez \thanks{Department of Mathematics, Universitat Politècnica de Catalunya, Barcelona, Spain. Centre de Recerca Matematica, Bellaterra,  Spain} \and Víctor Villegas-Morral \thanks{Department of Mathematics, Universitat Politècnica de Catalunya, Barcelona,  Spain.} \and Marie-Therese Wolfram \thanks{Mathematics Institute, University of Warwick, Coventry, UK.\\
G.E.R. and V.V.M are supported by the Spanish grant PID2022-143012NA-100 and G.E.R. is member of the Catalan research group 2021-SGR-00087.}}
\date{}

\begin{document}

\maketitle

\begin{abstract}
In this paper, we investigate the role of uninformed individuals in consensus formation within opinion-swarming models for self-propelled particles. The proposed models are inspired by empirical observations in animal swarming — particularly in schooling fish. We propose a coupled model that integrates spatial swarming dynamics with the evolution of individual opinions. Each individual is therefore described by its position, velocity, and a continuous opinion variable; it interacts through self-propulsion, alignment, attraction-repulsion forces, and opinion-based mechanisms. Building on classical bounded-confidence models, we introduce a three-population framework that distinguishes between leaders, followers, and uninformed individuals. Our analysis reveals that uninformed individuals, despite lacking any opinion bias, significantly influence group dynamics by diluting the effect of leaders and promoting more democratic decision-making. Numerical simulations demonstrate a variety of emergent behaviours, including flocking and milling. These findings support the role of uninformed agents in collective decision making and provide first analytical insights to understand leadership and opinion consensus in heterogeneous crowds. 
\end{abstract}

\section{Introduction}

Consensus formation in heterogeneous groups can be observed in many situations in our complex society as well as the life sciences; it is often linked to underlying structures, such as social networks, or spatio-temporal dynamics, as, for example in animal flocks. In this paper, we propose a coupled opinion-swarming model which is motivated by experiments of a heterogeneous group of strongly swarming fish conducted by Ian Couzin and co-workers, see \cite{couzin2011uninformed}. The experiments were motivated by the hypothesis that the presence of \emph{uninformed or na\"{i}ve individuals} destabilises the capacity for collective intelligence in groups \cite{conradt2009leading,mancur1968logic} and dilutes the role of leaders. Couzin et al. conducted experiments with golden shiners, a strongly schooling species of freshwater fish, to test this hypothesis. Their experiments were carried out to test the following hypothesis: a minority of opinionated leaders can dictate the group choice, however, the presence of uninformed or na\"{i}ve individuals inhibits the leaders' effect and restores control to the numerical majority. The opinion consensus in this case is not dictated by the leaders, but rather by the numerical majority.\\

There is an extensive literature on swarming models \cite{reynolds1987flocks,d2006self,degond2008continuum,dimarco2016self}, capable of reproducing the dynamics of large animal flocks, see for example \cite{vicsek2012collective} and the references therein. In particular, second order ODE systems \cite{cucker2007emergent,cucker2007mathematics}, in which the dynamics of each particle is described using Newton's law of motion, are among the most popular choices. In these models, pairwise interactions among individuals, such as attractive-repulsive potentials, lead to the formation of flocks or mills.
Works such as \cite{couzin2002collective} added some complexity to the above model, and in \cite{degond2008continuum} its macroscopic limit was derived.
The effect, emergence, and interactions of  follower leader populations have also been widely studied \cite{carrillo2010particle,fornasier2014mean,cai2023dynamic,cristiani2025kinetic,bernardi2021leadership} in different contexts.\\

Also, opinion formation models are well established in the applied mathematics and physics literature, see, for example, \cite{hegselmann2002opinion, krause2002living, deffuant2000, ben2005opinion, bordogna2007dynamic, motsch2014}. Bounded confidence models \cite{hegselmann2002opinion, krause2002living, lorenz2007} are among the most popular approaches, but also the effect of leaders, \cite{during2009, during2015opinion, albi2017recent} or external controls has been investigated \cite{qian2011,albi2016optimal, nugent2024}.\\

The proposed model is strongly motivated by experiments conducted by Couzin et al. \cite{couzin2011uninformed}, but similar dynamics have been observed in our society. The influence of leaders to control the public opinion is a concept that has been largely studied and goes back to \cite{taddicken2016people}, where it was found that, in the US presidential elections in 1940, interpersonal communication was more influential than media. An aspect in opinion formation that has been studied is the role of \emph{conviction} \cite{lasarsfeld1968peoples}, which can be interpreted as a measurable resistance to the change in opinion. For instance, in the case of the less active population, it allows them to partially align with the opinion of the leaders, while in the case of leaders the (strong) conviction will produce opinionated individuals. The ``conviction'' is also observed in social animals such as schooling fish, flocking birds, and herding ungulates where individual group members may be of low \emph{relatedness},  meaning that self-interests, i.e., strong convictions, play an important role in the decision making process \cite{krause2002living,olson2012logic}.
Therefore, it is well established that for both, animal groups and humans, the group decision can be influenced by a minority of leaders with strong convictions.\\

In this paper, we combine classical models in opinion dynamics, such as bounded confidence models \cite{hegselmann2002opinion}, with swarming models for self-propelled individuals \cite{d2006self}. The choice of the models introduced in Section \ref{sec: opinion swarming 1pop} is inspired by both empirical evidence and theoretical considerations from the study of biological swarming systems. In particular, \cite{d2006self} provides a foundational framework for understanding collective motion and emergent behaviour in systems composed of interacting agents. Their model incorporates soft-core pairwise interactions, self-propulsion, and velocity-dependent damping, providing a rich collection of dynamical behaviours such as flocking, vortex formation, and aggregation.

Incorporating such a swarming framework into an opinion dynamics model enables us to study not only how opinions evolve, but also how spatial proximity and physical interaction patterns affect consensus formation, leadership influence, and the role of uninformed individuals. 
To the best of our knowledge, the effects of uninformed individuals in consensus decision making has not yet been explored in the context of swarming mathematical modelling of opinion formation.

\paragraph{Main contributions of this paper:} 

\begin{enumerate}
    \item We propose and analyse an individual-based model in heterogeneous groups.
    \item We corroborate and extend our analytical findings with extensive computational experiments.
    \item Our computational experiments reproduce experimental findings for heterogeneous flocks of strongly herding fish, conducted by Couzin et al. in \cite{couzin2011uninformed}.
\end{enumerate}

\paragraph{Structure of the paper:} This paper is structured as follows: We start by sketching the experimental setup considered by Couzin et al. \cite{couzin2011uninformed} in Section \ref{sec:experiments}, which serves as a motivation of the mathematical model presented in Subsections \ref{sec:models} to \ref{sec: opinion only}. We then analyse the steady states for selected models and settings in Section \ref{sec:analysis}. In Section \ref{sec: numerical exp} we present extensive computational experiments, illustrating the rich dynamics of the proposed system and reproducing experimental findings.

\section{Mathematical models for opinion-swarming: motivation, assumptions and model development}\label{sec: opinion swarming 1pop}

In this section we  outline the experiments and findings presented by Couzin et al. in \cite{couzin2011uninformed}, which serve as a motivation of the proposed models presented in Subsections \ref{sec:models} to \ref{sec: opinion only}.

\subsection{Experimental setup}\label{sec:experiments}

Consider two groups of strongly schooling fish and let $N_L$ and $N_F$ (with $N_L < N_F$) denote the number fish in each group. We will refer to first group of $N_L$ fish as `leaders' and the second one of $N_F$ fish as `followers' for reasons outlined in the following. In the experiments, both groups were trained separately to move towards a yellow or blue target. The authors hypothesised that trained fish had an inherent preference for the colour yellow and that the additional training led to a strong preference for the yellow target in the first group. The second group of fish, trained to move towards the blue target, also developed a preference for the blue target. This preference was however, not as strong as the preference towards the yellow target in the first group. The authors therefore denoted the first group of fish (with the strong preference for yellow) as leaders, and the second group (with a weaker preference towards blue) as followers.\\

They then investigated the decision making process of a swarm that consists of $N_L$ `leader-type' fish, $N_F$ `follower-type' fish and $N_U$ untrained fish (with no preference for either target).\\
When only the two leader and follower populations were placed together, the leaders led the whole group towards the yellow target even though they were smaller in numbers (recall that $N_L < N_F$). However, when untrained fish were also present in the swarm, the swarm was more likely to move towards the blue target (returning the control to the numerical majority).\\
In the experiments performed in \cite{couzin2011uninformed}, the authors show the proportion of trials in which the blue target, which corresponds to the numerical majority, was chosen based on the number of untrained fish in the swarm, where each experiment was performed $18$ times.  They concluded that as the number of uninformed fish increased  ($N_U=0,\ 5,\ 10$), the school selects the majority preferred target more frequently. Couzin et al. \cite{couzin2011uninformed} also replicated the experimental results using a minimal mathematical model. Their computational results show that in case of a large number of uninformed fish, that is, $N_U\gtrsim 15$ for $N_L=5$ and $N_F=6$, the uninformed do not support the numerical majority.

\subsection{Mathematical models for opinion-swarming dynamics}\label{sec:models}

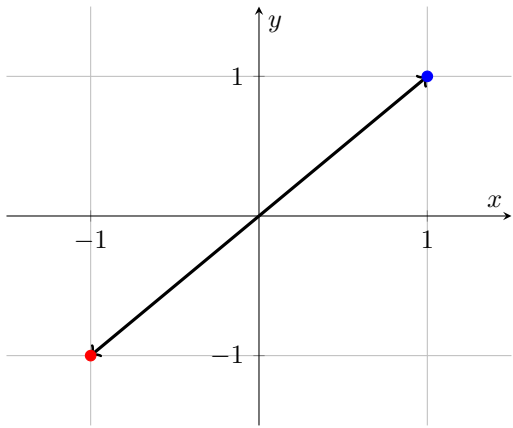
\begin{figure}
    \centering
    \begin{tikzpicture}
        \begin{axis}
            [
            axis lines=middle,
            width=0.5\linewidth,
            xlabel={$x$},
            ylabel={$y$},
            enlargelimits=0.25,
            xtick={0,1,-1},
            ytick={0,1,-1},
            grid=major
            ]
            \addplot[only marks, mark=*, color=blue] coordinates {(1,1)};
            \addplot[->, very thick] coordinates {(0,0) (1,1)};
            \addplot[only marks, mark=*, color=red] coordinates {(-1,-1)};
            \addplot[->, very thick] coordinates {(0,0) (-1,-1)};
        \end{axis}
    \end{tikzpicture}
    \caption{Blue and red targets in physical space. Individuals with opinion close to 1 have a stronger preference towards the top target (T, in blue), while individuals with opinion close to $-1$ prefer the bottom target (B, in red).}
    \label{fig:targets}
\end{figure}

Consider $N$ individuals, each characterised by the triple $(\bx_i, \bv_i, w_i)$, where $\bx_i\in \mathbb{R}^2$ corresponds to their position, $\bv_i \in \mathbb{R}^2$ corresponds to their velocity, and $w_i\in[-1,1]$ corresponds to their opinion. We assume there are two targets in the physical domain: a blue object at the top ($T$), and a red object at the bottom ($B$). Let $w_T=1$ and $w_B=-1$, where individuals with opinion close to $w_T$ have a stronger preference towards the top (blue) target, while individuals with opinion close to $w_B$ prefer to move towards the bottom (red) one, as seen in Figure \ref{fig:targets}.

\paragraph{Modelling assumptions}
\begin{enumerate}
    \item[A1.] Initial spatial positions are chosen uniformly in $\Omega=[-1,1]\times[-1,1]$.
    \item[A2.] Initial opinions are chosen uniformly in $[-1,1]$.
    \item[A3.] The spatial dynamics is driven by the following 
    \begin{itemize}
        \item[A3.1] Preferred speed: We assume that individuals wish to move at a preferred speed. This behaviour can be incorporated via a frictional force term of the form $(\alpha-\beta|\bv|^2)\bv$, for $\alpha,\ \beta\neq 0$ that describes this preferred speed, see \cite{d2006self} for further details.
        \item[A3.2] Attraction-repulsion forces: Depending on the distance between individuals, they can experience attraction (if the distance is larger than a certain value $\ell_a$), or repulsion (if the distance is shorter than $\ell_r$). Typical examples include the Morse-type potential $\mathcal{U}(|\bx|)$
        \begin{equation}\label{eq: morse potential}
            \mathcal{U}(z) = C_r e^{-\frac{z}{\ell_r}} - C_a e^{-\frac{z}{\ell_a}}\ ,
        \end{equation}
        where $C_a,C_r$ are the attractive and repulsive strengths, and $\ell_a,\ell_r$ are their respective length scales. However, different forms of interactions can also be considered.
        \item[A3.3] Velocity alignment due to opinions: Each individual adjusts their velocity to move towards the top or bottom target depending on their opinion $w_i$. If it is close to $w_T$ or $w_B$, that is $|w_s-w_i|<r_w$ for some $r_w \in \mathbb{R}^+$ and $s\in\{T, B \}$, then the individual will adjust its velocity to $\bv_T$ or $\bv_B$ respectively. Closeness can be enforced using the bounded confidence function $\psi(|w_s-w_i|)$, that is, 
        \begin{equation}\label{eq: bounded conf}
        \psi(|w_s-w_j|)=
            \begin{cases}
                1\ , \quad\textnormal{if} \quad|w_s-w_j|<r_w\ ,\\
                0\ , \quad\textnormal{otherwise}\ .
            \end{cases}
        \end{equation}
        Note that one could consider a modified version instead.
        \item[A4.] Compromise dynamics in opinion: Individuals align their opinions depending on whether they are sufficiently close in space and opinion. This can be included using a bounded confidence function that depends on both space and opinion, that is, 
        \begin{equation}\label{eq: bounded conf for x and w}
        \Phi(\bx,w)=\psi(|\bx_i-\bx_j|)\psi(|w_i-w_j|)=
            \begin{cases}
                1\, \quad\textnormal{if}\quad |w_i-w_j|<r_w \ \textnormal{and}\ |\bx_i-\bx_j|<r_\bx\ ,\\
                0\ , \quad \textnormal{otherwise}\ ,
            \end{cases}
        \end{equation}
        where $r_\bx$ is a spatial radius of interaction, and $r_w$ is the opinion radius. 
    \end{itemize}
\end{enumerate}

\subsection{Single population opinion-swarming dynamics}\label{sec: one population opinion-swarming}

We start by considering opinion-swarming dynamics for a single population. Individuals either have a preference to move to the top (T), and so $\bv_T = (1,1)$, or to the bottom (B), and so $\bv_B = (-1,-1)$. 
Based on the modelling assumptions (A3)-(A5), we obtain
\begin{subequations}\label{eq: one-population}
\begin{align}
    \dot{\bx}_i &= \bv_i\ ,\label{eq: position1} &&\\
    \dot{\bv}_i &= 
        (\alpha - \beta \lvert \bv_i \rvert^2) \bv_i 
        - \frac{1}{N} \sum_{j=1}^N \nabla_{\bx_i} \mathcal{U} (|\bx_j - \bx_i|) 
        + \sum_{s=T,B}{\gamma_s\,\psi(|w_s-w_i|)(\bv_s-\bv_i)}\ ,\label{eq: velocity1}\\
    \dot{w}_i &=
        \frac{1}{N}\sum_{j=1}^N\Phi_{ij}(\bx,w)(w_j-w_i)
        +\sum_{s=T,B}{{\tau_s}(w_s-w_i)}\ ,\label{eq: opinion1}
\end{align}
\end{subequations}
where we choose $\mathcal{U}$ as in \eqref{eq: morse potential}, $\alpha, \beta, \gamma_s, \tau_s \in \mathbb{R}^+$
and $\Phi$ is defined as in \eqref{eq: bounded conf for x and w}. 

We recall that the first term in \eqref{eq: velocity1} accounts for adjustment to move at the desired speed (see Assumption (A3.1)), the second for attractive-repulsive interactions among individuals (see Assumption (A3.2)), and the third for the alignment in velocity of an individual based on its preference (see Assumption (A3.3)). Note that, if $|w_s-w_i|>r_w$, we say that individual $i$ is ``undecided'', namely, its velocity will only depend on the flocking velocity, but not on the preference.

Equation \eqref{eq: opinion1} describes the evolution of opinion/preference -- the first term corresponds to alignment (also known as consensus formation) if individuals are sufficiently close in space and opinion (see Assumption (A4)). The second term allows including inherent preferences to the top or bottom target. Note that we assume that both parameters $\gamma_s$ and $\tau_s$ are constant in time; however, in general and depending on the individuals, parameters such as conviction could change over time.

\begin{remark}
Variants of \eqref{eq: one-population} could also be considered. For instance, we can describe a model with no ``undecided'' individuals in which the velocity evolves according to 
\begin{equation*}
    \dot{\bv}_i =
        (\alpha - \beta \lvert \bv_i \rvert^2) \bv_i
        - \frac{1}{N} \sum_{j=1}^N \nabla_{\bx_i} \mathcal{U} (|\bx_j - \bx_i|)
        + \gamma_B\mathds{1}_{w_i< 0}(\bv_B-\bv_i)
        + \gamma_T\mathds{1}_{w_i> 0}(\bv_T-\bv_i)\ .
\end{equation*}
The resulting dynamics are very similar to those obtained with \eqref{eq: one-population}. 
\end{remark}

\begin{remark}
Couzin et al. proposed a discrete in time model in \cite{couzin2002collective}, which describes the spatial evolution of individuals in the case of attractive-repulsive interactions and a preferred direction (which depends on the type of fish). Furthermore, they add a stochastic component to account for random influences during motion, which affected the rotational angle in their model.
\end{remark}

\subsection{Three population opinion-swarming dynamics}\label{sec:threepopmodel}

In this section we generalise to multiple populations the opinion-swarming model introduced in Section  \ref{sec: one population opinion-swarming}. Let $\mathcal{G}$ denote the set of different population groups. Then, the position, velocity, and opinion of an individual $i$ in a population  $p\in \mathcal{G}$ of size $N_p$ evolves according to
\begin{subequations}\label{eq: opinion_sawarm 3pop}
    \begin{align}
        \dot{\mathbf{x}}_i&=
            \mathbf{v}_i\ ,\\
        \dot{\mathbf{v}}_i&=
            (\alpha-\beta|\mathbf{v}_i|^2)\mathbf{v}_i
            -\sum_{q\in\mathcal{G}}\frac{u_{pq}}{N_{q}}\sum_{j\in q} 
            \nabla_{\mathbf{x}_i}\mathcal{U}_{pq}(|\mathbf{x}_j-\mathbf{x}_i|)
            +\sum_{s=T,B} {\gamma_s^{p}\,\psi(|w_s-w_i|)(\bv_s
            -\bv_i)}\ ,\\
        \dot{w}_i&=
            \sum_{q\in\mathcal{G}}\frac{k_{pq}}{N_{q}}\sum_{j\in q} \Phi_{ij}(\bx,w)(w_j-w_i)
            +\sum_{s=T, B} {\tau_s^{p}(w_s-w_i)}\ , \label{eq:swarming-w}
    \end{align}
\end{subequations}
where $\Phi$ is defined as in \eqref{eq: bounded conf for x and w} and the potentials $\mathcal{U}_{pq}$ can be different interacting potentials depending on the populations. We will, however set $\mathcal{U}_{pq} = \mathcal{U}$ for simplicity in the following. The constants $u_{pq}$ and $k_{pq} \in \mathbb{R}^+$ describe the strength of the spatial interactions and opinions. The parameters that control the alignment towards preferred targets are denoted as $\gamma_s^{p}$ for velocities and $\tau_s^{p}$ for opinions where $p\in\mathcal{G}$. 

Motivated by the experiments outlined in Section \ref{sec:experiments}, we consider three populations: leaders (L), followers (F), and uninformed individuals (U). Therefore, $\mathcal{G}=\{L,F,U\}$ and $N=N_L+N_F+N_U$. The preferences and parameters for each group can be summarised as follows:

\begin{enumerate}
    \item[L.] Leaders exhibit a strong preference for the top target, therefore $\tau_T^L > 0$ for $w_T=1$ and $\mathbf{v}_T=(1,1)$. Conversely, they have no preference for the bottom target, and hence we set $\tau_B^L=0$.
    \item[F.] Followers have a weaker preference for the bottom target. Hence, $\tau_B^F > 0$ with $\tau_B^F < \tau_T^L$,  driving them towards $\bv_B=(-1,-1)$. Moreover, they have no preference for the top target, therefore $\tau_T^F = 0$.
   
    \item[U.] Uninformed individuals have no inherent preference for either target. Their position and velocity change solely due to  interactions with individuals within their own group and with members of other groups. Consequently, we set $\tau_T^U = \tau_B^U = 0$. 
\end{enumerate}

To simplify notation we define $\tau_T\coloneqq\tau_T^L$ and $\tau_B\coloneqq\tau_B^F$ and set $\gamma_s^L = \gamma_s^F = 1$.

\subsection{Space-independent model}\label{sec: opinion only}

We conclude by considering the opinion dynamics among the three groups (omitting the spatial dependence). Then \eqref{eq: opinion_sawarm 3pop} reduces to \eqref{eq:swarming-w}, in particular
\begin{subequations}\label{eq:opinion-only}
\begin{align}
    \text{if }i\in L,\hspace{4em}\dot{w}_i&=
        \sum_{q\in\mathcal{G}}\frac{k_{Lq}}{N_{q}}\sum_{j\in q} {\psi(|w_i- w_j|)(w_j-w_i)}
        +{\tau_T(w_T-w_i)}\ , \label{eq: leaders}\\
    \text{if }i\in F,\hspace{4em}\dot{w}_i&=
        \sum_{q\in\mathcal{G}}\frac{k_{Fq}}{N_{q}}\sum_{j\in q} {\psi(|w_i- w_j|)(w_j-w_i)}
        +{\tau_B(w_B-w_i)}\ , \label{eq: follow}\\
    \text{if }i\in U,\hspace{4em}\dot{w}_i&=
        \sum_{q\in\mathcal{G}}\frac{k_{Uq}}{N_{q}}\sum_{j\in q} {\psi(|w_i- w_j|)(w_j-w_i)}\ , \label{eq: uninformed}
\end{align}
\end{subequations}
where $\psi(|w_i- w_j|)$ is given by \eqref{eq: bounded conf}. We recall that $k_{Lq}, k_{Fq}$, and $k_{Uq}$ are positive constants, and that $\tau_T \gg \tau_B$.\\
System \eqref{eq:opinion-only} is more amenable to mathematical analysis, as we will see in Section \ref{sec:analysis}. In particular, we will be able to calculate its stationary states and analyse the effect of the modelling parameters.

\section{Large time behaviour and stationary states in simplified settings}\label{sec:analysis}
In this section, we present some analytical results for the previously discussed systems. These results help validate the accuracy of the numerical simulations in Section \ref{sec: numerical exp} for general choices of parameters, and provide deeper insights into the structure and behaviour of the equations.

\subsection{Space-independent model}

Here, we present an analytical solution for the mean opinion of each population. We start by considering the space-independent system \eqref{eq:opinion-only} with $N_U=0$, assuming that all individuals interact, hence $\psi\equiv1$, and that all populations interact, hence $k_{pq}=1$ for $p,q \in \lbrace L, F \rbrace$. The mean opinion, $\bar w=\frac{1}{N}\sum_{j\in p}w_j$, satisfies
\begin{subequations}\label{eq:opinion-only-sum-equations}
\begin{align}
    \dot{\bar{w}}_L &= \bar{w}_L - (1+\tau_T)\bar{w}_L  + \tau_T w_T\ , \label{eq:mean-leaders}\\
    \dot{\bar{w}}_F &= \bar{w}_F - (1+\tau_B)\bar{w}_F + \tau_B w_B\ .
\end{align}
\end{subequations}
In order to compute a solution for $\bar w_p$, we first solve the initial value problem \eqref{eq:opinion-only-sum-equations} for $\bar w_p$ subject to some initial conditions given by $\ \bar w_p(0)=\bar w_{p_0}$.

Setting $\ a=1+\tau_T,\ b=1+\tau_B,\ f=\tau_T w_T,$ and $g=\tau_B w_B$, we write \eqref{eq:opinion-only-sum-equations} in matrix form 
$\dot{\mathbf{z}} = \mathbf{A}\mathbf{z} + \mathbf{u}$, or equivalently,
\begin{equation}\label{eq:opinion-only-sum-system}
    \begin{pmatrix}
        \dot{\bar{w}}_L \\
        \dot{\bar{w}}_F
    \end{pmatrix}
    =
    \begin{pmatrix}
        -a & 1 \\
        1 & -b
    \end{pmatrix}\begin{pmatrix}
         \bar{w}_L \\
         \bar{w}_F
    \end{pmatrix}
    +
    \begin{pmatrix}
        f \\
        g
    \end{pmatrix}\ .
\end{equation}
The matrix $\mathbf{A}$ has eigenvalues
$$\lambda_{1}=\frac{1}{2}\left(-a-b+ \sqrt{(a-b)^2+4}\right) \ , \qquad \lambda_{2}=\frac{1}{2}\left(-a-b- \sqrt{(a-b)^2+4}\right)\ ,$$
% \lambda_{\pm}=\frac{1}{2}\left(-a-b\pm \sqrt{(a-b)^2+4}\right)
and eigenvectors $\mathbf{y}_i = (1,\ \lambda_i + a)^T$. The general solution $\mathbf{z}(t)$ of the system is given by combining the homogeneous solution $\mathbf{z}^h$ for $\mathbf{u} = (f,g)^{T}=0$,
\begin{equation}
    \mathbf{z}^h(t)
    =\begin{pmatrix}
        \bar{w}_L^h \\
        \bar{w}_F^h
    \end{pmatrix}
    = c_1 e^{\lambda_1 t}\mathbf{y}_1 + c_2 e^{\lambda_2 t}\mathbf{y}_2 \ ,
\end{equation}
and the equilibrium solution given by
\begin{equation}\label{eq:opinon-only-sum-equilibrium}
    \mathbf{z}^*
    =\begin{pmatrix}
        \bar{w}_L^* \\
        \bar{w}_F^*
    \end{pmatrix}
    = -\mathbf{A}^{-1}\mathbf{u}
    = \frac{1}{ab-1}
    \begin{pmatrix}
        g + bf \\
        f + ag
    \end{pmatrix} \ .
\end{equation}
Therefore, the general solution is
\begin{equation}\label{eq:opinon-only-sum-solution}
    \mathbf{z}(t) = \mathbf{z}^h + \mathbf{z}^* = c_1 e^{\lambda_1 t}\mathbf{y}_1 + c_2 e^{\lambda_2 t}\mathbf{y}_2 + \mathbf{z}^*\ ,
\end{equation}
where
\begin{equation*}
    c_2 = \frac{(\bar{w}_{F_0}-\bar{w}_F^*)-(\bar{w}_{L_0}-\bar{w}_L^*)(\lambda_1+a)}{\lambda_2-\lambda_1}\ ,
    \qquad 
    c_1 = (\bar{w}_{L_0}-\bar{w}_L^*) - c_2\ ,
\end{equation*}
are constants that depend on the initial opinions. Note that the values of $\tau_T,\tau_B$ considered satisfy $ab-1\neq0$, and that $\lim_{t\rightarrow\infty}{\mathbf{z}(t)}=\mathbf{z}^*$. \\

\noindent Now, consider the case where we include uninformed individuals ($N_U>0$) such that these remain uninformed for all times and only followers are affected by them ($k_{Up}=0 ,\ k_{FU}=1$). In this system, $\dot{\bar{w}}_L$ is still given by \eqref{eq:mean-leaders}, and $\dot {\bar{w}}_F$ becomes
\begin{equation*}
    \dot{\bar{w}}_F = \bar{w}_F - (2+\tau_B)\bar{w}_F + \tau_B w_B + \bar{w}_{U_0}\ ,
\end{equation*}
where the mean opinion of the uninformed is constant in time, i.e., $\bar{w}_U\equiv \bar{w}_{U_0}$. Solution \eqref{eq:opinon-only-sum-solution} therefore holds by changing parameters $b$ and $g$ to $\tilde{b}=2+\tau_B$ and $\tilde{g}=\tau_B w_B + \bar{w}_{U_0}$.\\

Let us consider that all the uninformed individuals' opinions are zero, and thus $\bar{w}_{U_0}=0$. From \eqref{eq:opinon-only-sum-equilibrium}, substituting $f$ and $g$, we observe that the steady state of the mean opinion of both followers and leaders depends only on reference parameters $w_s,\ \tau_s$, and not on population sizes or initial opinions:
\begin{equation}\label{eq:opinon-only-sum-equilibrium-2}
    \mathbf{z}^*
    =\begin{pmatrix}
        \bar{w}_L^* \\
        \bar{w}_F^*
    \end{pmatrix}
    = \frac{1}{a\tilde{b}-1}
    \begin{pmatrix}
        \tilde{b}\tau_T w_T + \tau_B w_B \\
        \tau_T w_T + a\tau_B w_B
    \end{pmatrix} \ .
\end{equation}
The analytical steady state solutions validate the numerical results of the simulations as we discuss in Section \ref{sec: numerical exp} for the case $\psi\equiv1$, i.e., all individuals interact.

\subsubsection{Energy}\label{sec:energy}

We define, for both $N_U=0$ and $N_U>0$, the energy as 
\begin{equation}
    \mathcal{E}(t)=\frac{1}{2N}\left(
        \sum_{i\in L}(w_i-\bar{w}_L^*)^2+\sum_{i\in F}(w_i-\bar{w}_F^*)^2
    \right)\ .
\end{equation}

\begin{figure}[ht]
        \centering
        \begin{subfigure}{0.35\linewidth}
            \centering
            \includegraphics[height=4cm]{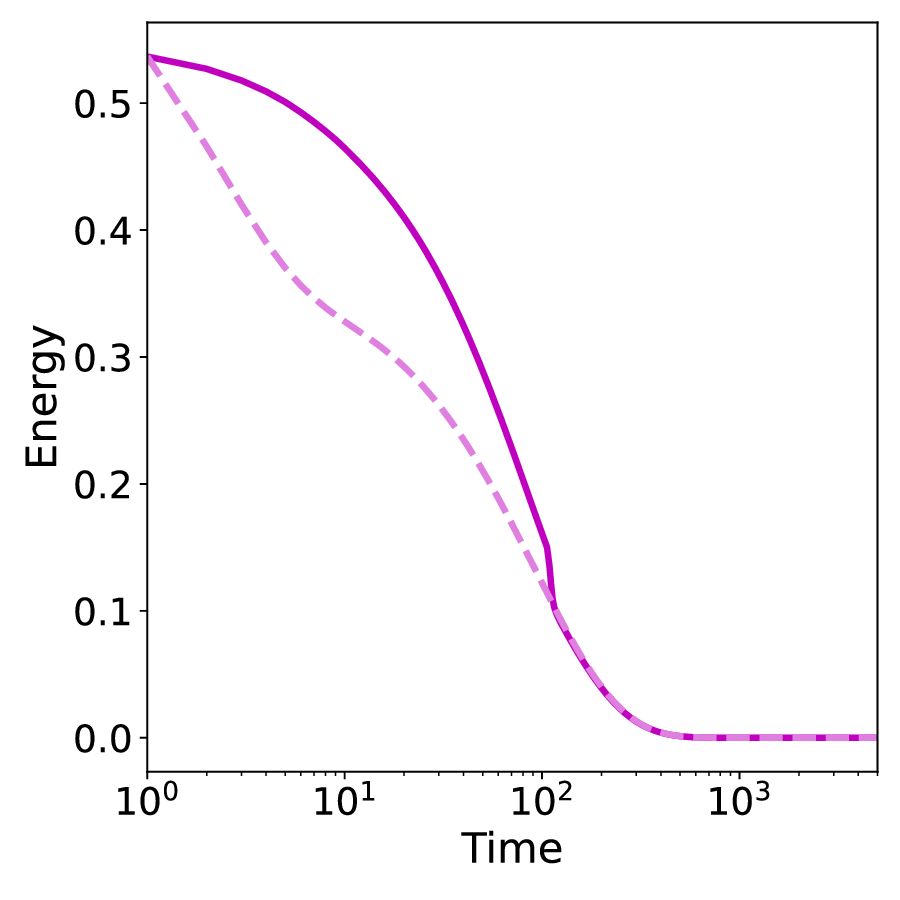}
            \caption{}
        \end{subfigure}
        \begin{subfigure}{0.35\linewidth}
            \centering
            \includegraphics[height=4cm]{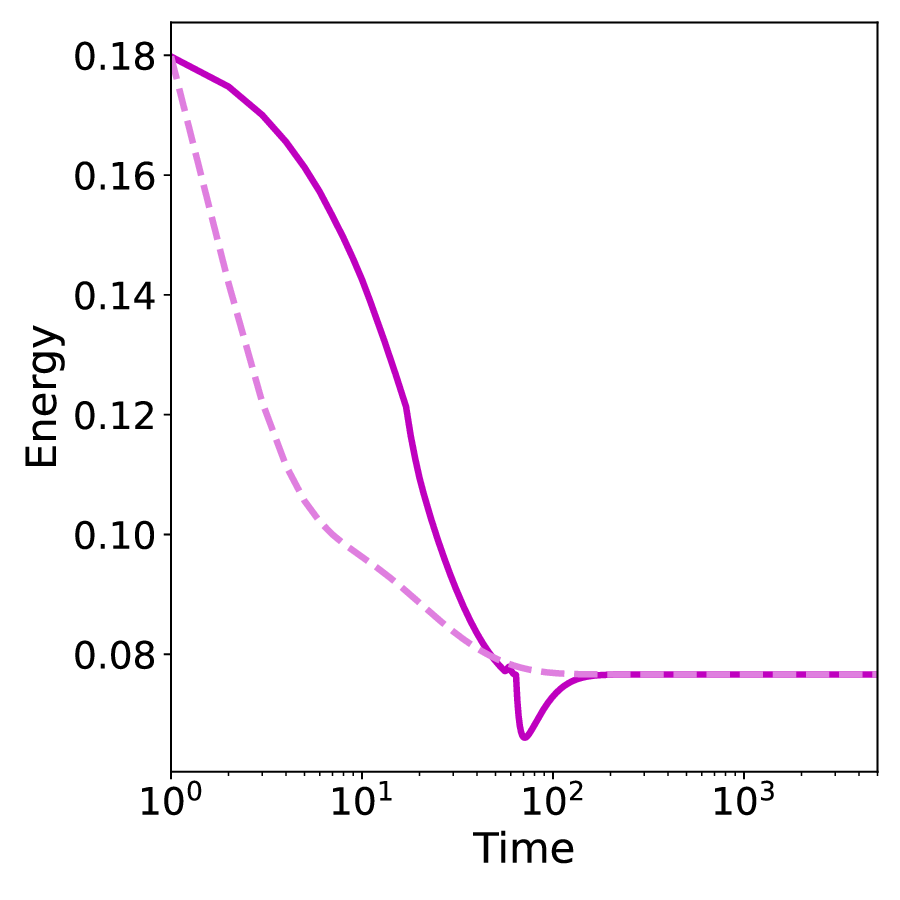}
            \caption{}
        \end{subfigure}
        \caption{(a) Energy for $N_L=5$, $N_U=6$, $N_U=0$. (b) Energy for $N_L=5$, $N_U=6$, $N_U=10$. The continuous line represents the numerical solution for $r_w=0.5$, and the dashed line the analytic solution for $\psi\equiv1$.}\label{fig: energy}
\end{figure}
As we observe in Figure \ref{fig: energy}, the energy decreases for all time and stays constant when the system achieved consensus at around $t=10^3$ for the case of $N_U=0$, and around $t=10^2$ for the case with $N_U=10$ (see Figure \ref{fig:opinions-only-no-uninformed-analytical} for the opinion's evolution). The decay of the energy in case of no uninformed individuals is further supported by the following calculation
\begin{align*}
    \frac{d}{dt} \mathcal{E}(t) = -(\bar{w}_L - \bar{w}_F)^2 - \tau_T \left(\frac{1}{N}\sum_{i\in L} w_i - \bar{w}^*_T\right)^2 - \tau_B \left(\frac{1}{N}\sum_{i \in F} w_i - \bar{w}^*_B\right)^2 + \frac{4\tau_T \tau_F}{\tau_T + \tau_B + \tau_T \tau_B},
\end{align*}
which can be obtained after some tedious computations. We see that the first three terms have a negative sign, the only positive contribution comes from the last term. This term is, however, of much smaller magnitude and therefore the first three terms most likely dominate the dynamics.

\subsection{Opinion-swarming model}\label{sec: opinion-swarming analytic}

The derivation of an analytic solution for the case of the system \eqref{eq: one-population} is not straightforward since there are non-linearities coming from the term $(\alpha-\beta|\bv_i|^2)\bv_i$. We should also reconsider the assumption $\psi\equiv 1$, which removes all dependence of the velocity on the opinion variable. Therefore, for this model we are going to focus only on the steady state.

Let us recall the equations that describe the opinion-swarming dynamics for an arbitrary population $p$,
\begin{subequations}
\begin{align}
    \dot{\mathbf{x}}_i&=
        \mathbf{v}_i\ ,\\
    \dot{\mathbf{v}}_i&=
        (\alpha-\beta|\mathbf{v}_i|^2)\mathbf{v}_i
        -\sum_{q\in\mathcal{G}}\frac{u_{pq}}{N_{q}}\sum_{j\in q} 
        \nabla_{\mathbf{x}_i}\mathcal{U}_{pq}(|\mathbf{x}_j-\mathbf{x}_i|)
        +\sum_{s=T,B} {\gamma_s^{p}\,\psi(|w_s-w_i|)(\bv_s
        -\bv_i)}\ \label{eq:swarming-v},\\
    \dot{w}_i&=
        \sum_{q\in\mathcal{G}}\frac{k_{pq}}{N_{q}}\sum_{j\in q} \Phi_{ij}(\bx,w)(w_j-w_i)
        +\sum_{s=T, B} {\tau_s^{p}(w_s-w_i)}\ .
\end{align}
\end{subequations}

There are several possible scenarios for steady state solutions, in particular we consider a \emph{cohesive motion} steady state for $\dot\bv$ since it is the most relevant in Section \ref{sec: numerical exp}. In this case, all individuals in each population $p$ move with the same constant velocity $\bv^*_p$.
Assuming $\bv_{i} = \bv^*_p \neq 0$ for all $i$ in $p$, equation \eqref{eq:swarming-v} becomes
\begin{equation*}
    0 = (\alpha - \beta |\bv^*_p|^2) \bv^*_p -
    \sum_{q\in \mathcal{G}}\frac{u_{pq}}{N_q}
    \sum_{j\in q} \nabla_{\bx_{i}} \mathcal{U}_{pq} (|\bx_{i} - \bx_{j}|)
    + \sum_{s=T,B} \gamma_s^p \,\psi(|w_s-w_i|) (\bv_s - \bv^*_p)\ .
\end{equation*}
Since $\bv_{i} = \bv^*_p$ for all $i\in p$, then $\bx_{i}(t)=\bv^*_pt$ for all $i\in p$. Therefore, 
$\mathcal{U}_{pp} (|\bx_{i} - \bx_{j}|)=0,$
which yields, for each population
\begin{align}
    \bv_L^*&=\frac{\sum\limits_{q=F,U}\frac{u_{Lq}}{N_q}\sum\limits_{j\in q}\nabla_{\bx_i} \mathcal{U}_{Lq}(|\bx_i-\bx_j|)-\sum\limits_{s=T,B}\gamma_s^L\,\psi(|w_s-w^*_L|)\bv_s}{\alpha-\beta|\bv^*_L|^2-\sum\limits_{s=T,B}\gamma_s^L\,\psi(|w_s-w^*_L)}\ ,\\
    \vspace{0.1cm}
    \bv_F^*&=\frac{\sum\limits_{q=L,U}\frac{u_{Fq}}{N_q}\sum\limits_{j\in q}\nabla_{\bx_i} \mathcal{U}_{Fq}(|\bx_i-\bx_j|)-\sum\limits_{s=T,B}\gamma_s^F\,\psi(|w_s-w^*_F|)\bv_s}{\alpha-\beta|\bv^*_F|^2-\sum\limits_{s=T,B}\gamma_s^F\,\psi(|w_s-w^*_F|)}\ ,\\
    \vspace{0.1cm}
    \bv_U^*&=\frac{\sum\limits_{q=L,F}\frac{u_{Uq}}{N_U}\sum\limits_{j\in q}\nabla_{\bx_i}\mathcal{U}_{Uq}(|\bx_i-\bx_j|)}{\alpha-\beta|\bv_U^*|^2}\ ,
\end{align}
where $w_L^*$ and $w_F^*$ are the opinion steady states for leaders and followers respectively. 

\begin{remark}\label{rem: steady state velo}
    For the case where we have a unique steady state for the velocities, $\bv=\bv^*$ for all populations (or equivalently, for the one population case), then $\mathcal{U}_{pq}(|\bx_i-\bx_j|)=0$ and
    \begin{equation}
        \bv^*=\frac{\sum_{s=T,B} \gamma_s\, \psi(|w_{s}-w^*|)\bv_s}{\sum_{s=T,B} \gamma_s\, \psi(|w_{s}-w^*|)- (\alpha - \beta |\bv^*|^2)}\ .\label{eq: one velocity steady state}
    \end{equation}
    If the steady state opinion $w^*$ is close to $1$, then $\sum_{s=T,B} \gamma_s\, \psi(|w_{s}-w^*|)=\gamma_T$, therefore
    \[
    \bv^*=\frac{\gamma_T\bv_T}{\gamma_T-(\alpha-\beta|\bv^*|^2)}\ .
    \]
    Since in all numerical examples we consider $\gamma=1$ and $\alpha=1$, we finally get $
    \bv^*=\frac{\bv_T}{\beta|\bv^*|^2}$,
    which means that the steady state velocity has the same direction as $\bv_T=(1,1)$. Similarly, if the steady state opinion $w^*$ is close to $-1$, we have  $\bv^*=\frac{\bv_B}{\beta|\bv^*|^2}$.
    
    In the particular case when $|\bv^*|=\sqrt{\frac{\alpha}{\beta}}$ then
    $\sum_{s=T,B} \gamma_s \psi(|w_{s}-w^*|)(\bv_s - \bv^*) = 0\ .$
\end{remark}

The condition for the opinion steady state is
\begin{equation}
    \sum_{q\in \mathcal{G}}\frac{k_{pq}}{N_q}
    \sum_{j\in q} \Phi_{ij}(\bx,w) (w_{j} - w_{i})
    = - \sum_{s=T,B} \tau_{s}^p (w_{s} - w_i)\ .
\end{equation}
For $\Phi_{ij}(\bx,w)\equiv 1$ and for one population we recover the steady states derived in \eqref{eq:opinon-only-sum-equilibrium}.

\section{Numerical experiments}\label{sec: numerical exp}

We conclude by illustrating the behaviour of the proposed models with extensive computational results. We start with the opinion model in Section \ref{sec: numerics opinion only} before discussing more complex dynamics in the case of single and multi-species swarming-opinion formation models.
All simulations were carried out using Python libraries \cite{simulation_repo}
 with time steps of size $0.1$, for the space-independent model, and $0.01$ for the opinion-swarming case.

\subsection{Space-independent model}\label{sec: numerics opinion only}

In this section we consider \eqref{eq:opinion-only}. Based on Couzin's experiments we set $N_L=5$, $N_F=6$ and $N_U=10$, as well as $\tau_T=0.1$ for the leaders and $\tau_B=0.01$ for the followers (to account for the leaders' stronger conviction to go to the top target). Furthermore, we set the opinion radius to $r_w=1$. The initial opinions of the followers and leaders are chosen uniformly at random from $[-1,1]$. The initial opinion of uninformed is zero.

\noindent Throughout this section, we plot the dynamics on logarithmic scales for better presentation. In the following, we focus on three specific cases:
 \begin{itemize}
    \item Example 1: only leaders and followers interact, that is,  $k_{UL}=k_{LU}=k_{UF}=k_{FU}=0$, and $k_{pq}=1$ otherwise.
    \item Example 2: all individuals interact, therefore $k_{pq} = 1$ for all $p, q$.
    \item Example 3: uninformed stay uninformed (their dynamics are not influenced by leaders, followers, or other uninformed, yet they influence followers). This implies  
    \begin{align*}
        k_{UU}=k_{UF}=k_{UL}=0 \text{ and }    k_{LU}=0,\ k_{FU}=1.    
    \end{align*}
    Note that this setting is closest to the experimental conditions, since it has not been observed that uninformed fish develop a preference towards either target in time.
\end{itemize}

\begin{figure}[!ht]
    \centering
    \begin{subfigure}{0.25\textwidth}
        \centering
        \includegraphics[height=4.25cm]{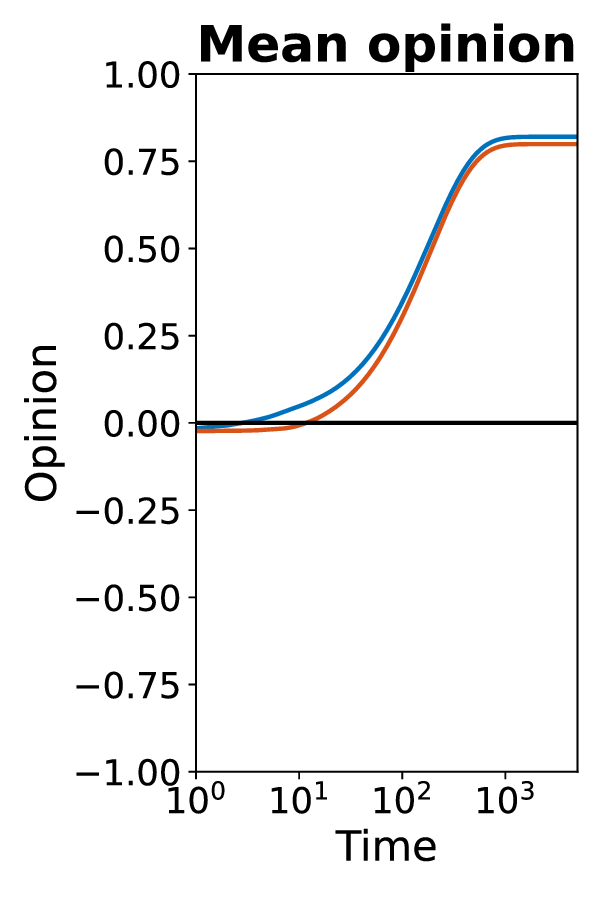}
        \caption{}\label{fig:opinion-only-means-a}
    \end{subfigure}
    % % \hspace{0.005\textwidth}
    \begin{subfigure}{0.25\linewidth}
        \centering
        \includegraphics[height=4.25cm]{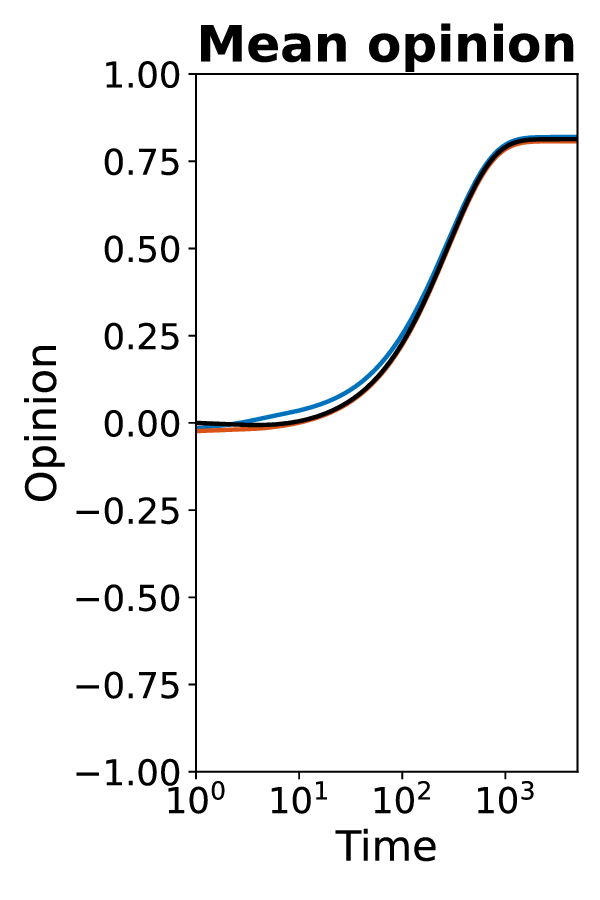}
        \caption{}\label{fig:opinion-only-means-b}
    \end{subfigure}
    % \hspace{0.005\textwidth}
    \begin{subfigure}{0.25\linewidth}
        \centering
        \includegraphics[height=4.25cm]{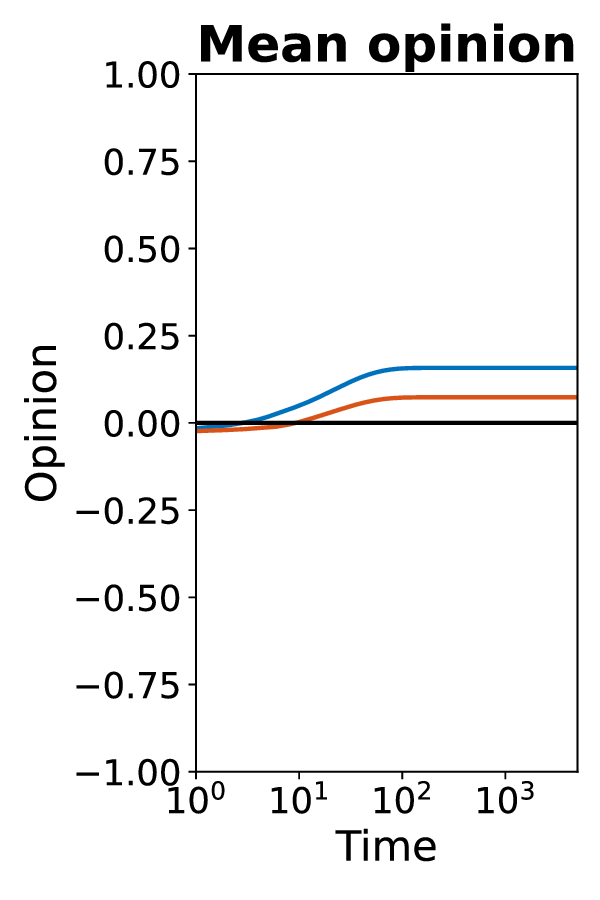}
        \caption{}\label{fig:opinion-only-means-c}
    \end{subfigure}
    \caption{
        Averages opinion over 100 different initial conditions for the followers (blue), leaders (red), and uninformed (black), corresponding to Examples 1, 2 and 3, respectively. 
    }
    \label{fig:opinion-only-means}
\end{figure}

Figure \ref{fig:opinion-only-means} shows the mean opinion of $100$  different initial conditions for the followers and leaders for Examples $1$-$3$. The mean opinion is computed by first computing the mean of each population over time, and then averaging this mean among all initial conditions.
In Figure \ref{fig:opinion-only-means-a}  we observe how the leaders' opinion dominates and the overall opinion approaches $w_T=1$. For Example 2 we observe in Figure \ref{fig:opinion-only-means-b} that the whole population tends towards the leaders' preferred opinion. The uninformed act as followers, ultimately converging towards the final consensus shaped by the leaders.
In Example 3, as presented in Figure \ref{fig:opinion-only-means-c}, we see the influence of the uninformed. While in  Example 1 (Figure \ref{fig:opinion-only-means-a}) the final average opinion is approximately $0.8$,  in the case of Figure \ref{fig:opinion-only-means-c} the averaged opinion is approximately $0.3$. Therefore, we observe that, even when the uninformed individuals remain uninformed and interact only with the followers, the final average opinion tends to align more closely with the majority preference, i.e., the followers' preference towards $w_B=-1$.

\begin{figure}[!ht]
    \centering
    \begin{subfigure}{0.36\linewidth}
        \centering
        \includegraphics[height=4cm]{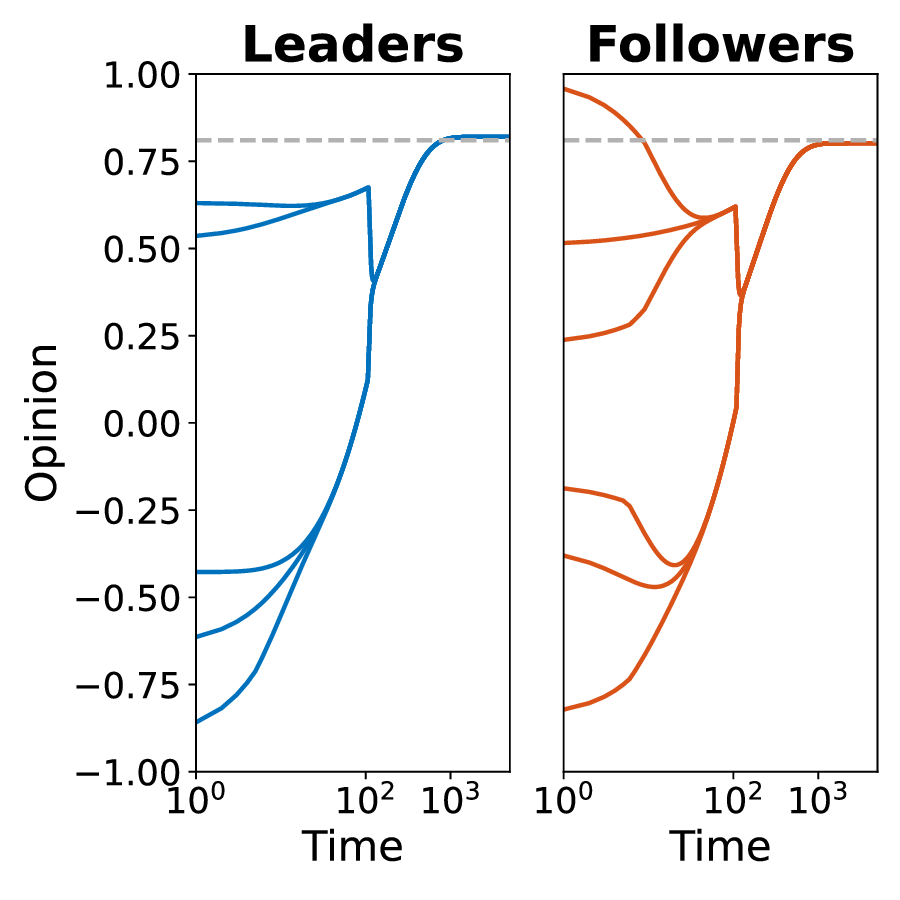}
        \caption{Evolution for $r_w=0.5$.}\label{fig: numerical-rw-05}
    \end{subfigure}
    \begin{subfigure}{0.30\linewidth}
        \centering
        \includegraphics[height=4cm]{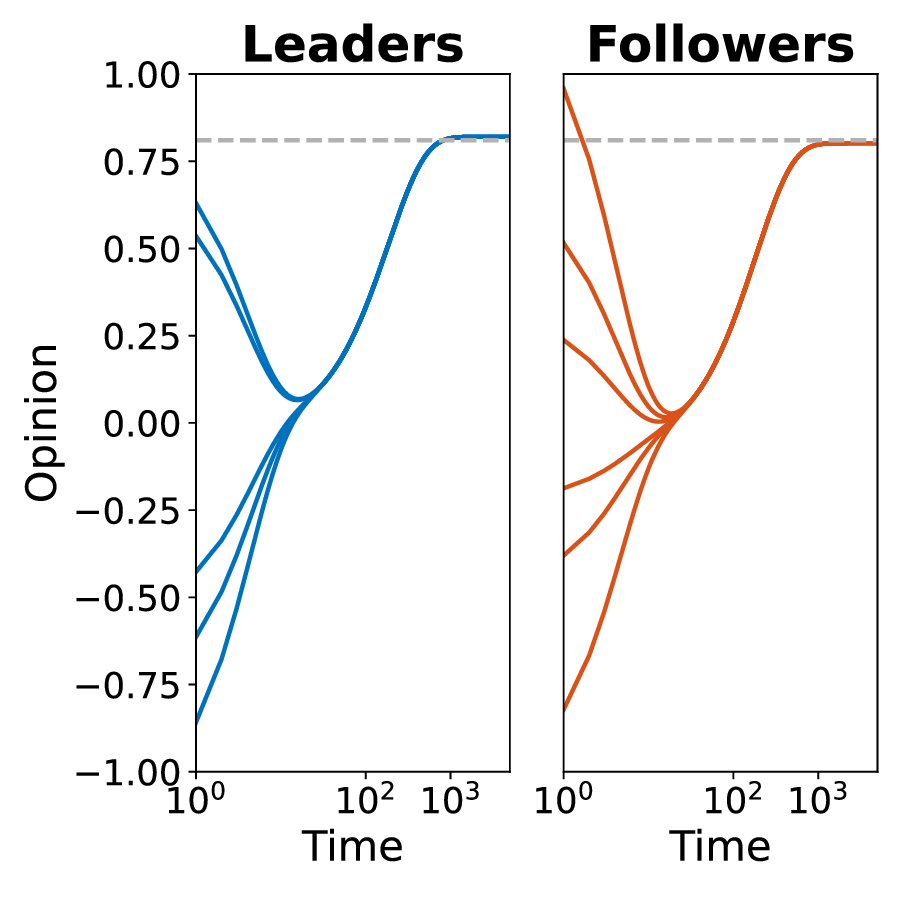}
        \caption{Evolution for $\psi\equiv 1$.}\label{fig: numerical-no-rw}
    \end{subfigure}
    \begin{subfigure}{0.30\linewidth}
        \centering
        \includegraphics[height=4cm]{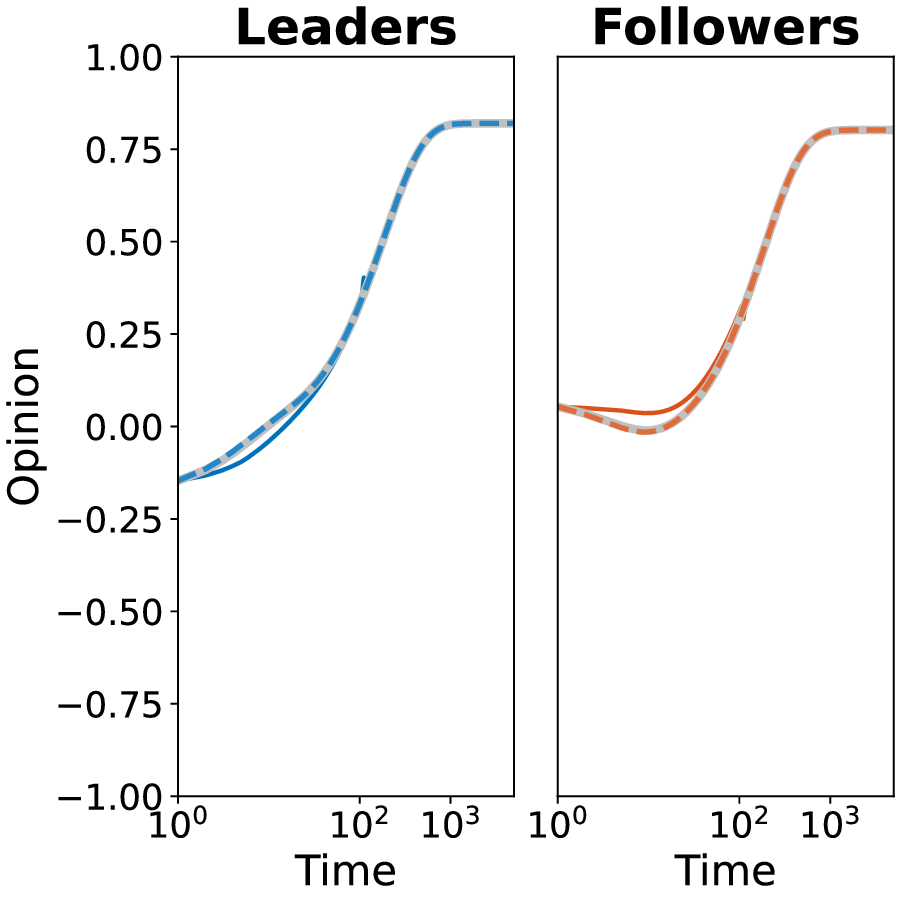}
        \caption{Mean opinion.}\label{fig: comparison-nouninformed}
    \end{subfigure}
    
    \vspace{1em}
    
    \begin{subfigure}{0.36\linewidth}
        \centering
        \includegraphics[height=4cm]{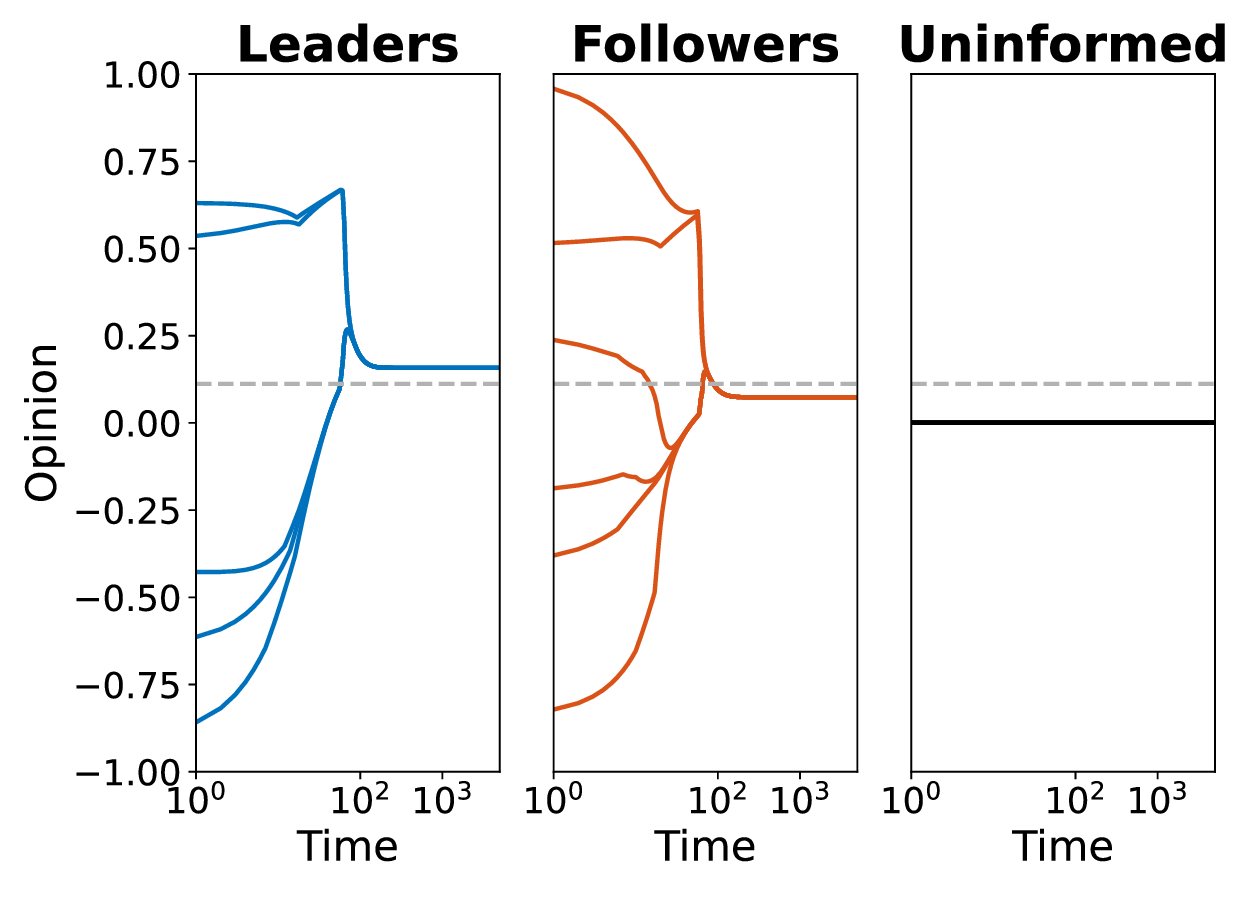}
        \caption{Evolution for $r_w=0.5$.}\label{fig: numerical-rw-05-uninformed}
    \end{subfigure}
    \begin{subfigure}{0.30\linewidth}
        \centering
        \includegraphics[height=4cm]{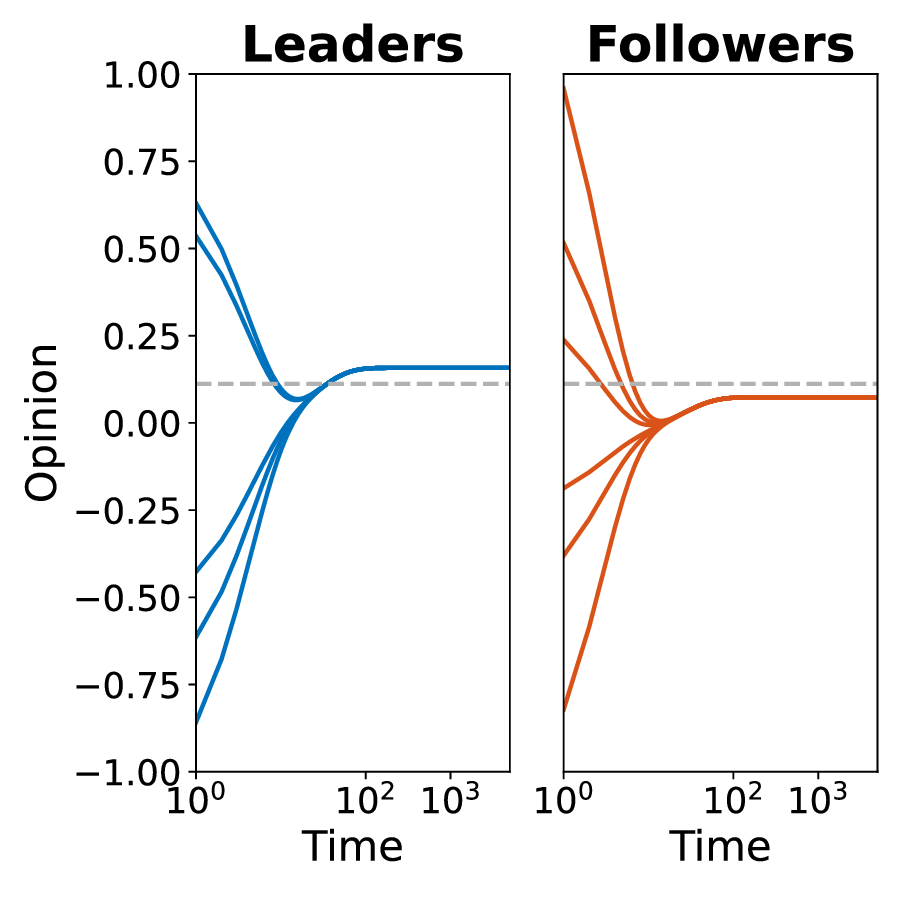}
        \caption{Evolution for  $\psi\equiv 1$.}\label{fig: numerical-no-rw-uninformed}
    \end{subfigure}
    \begin{subfigure}{0.30\linewidth}
        \centering
        \includegraphics[height=4cm]{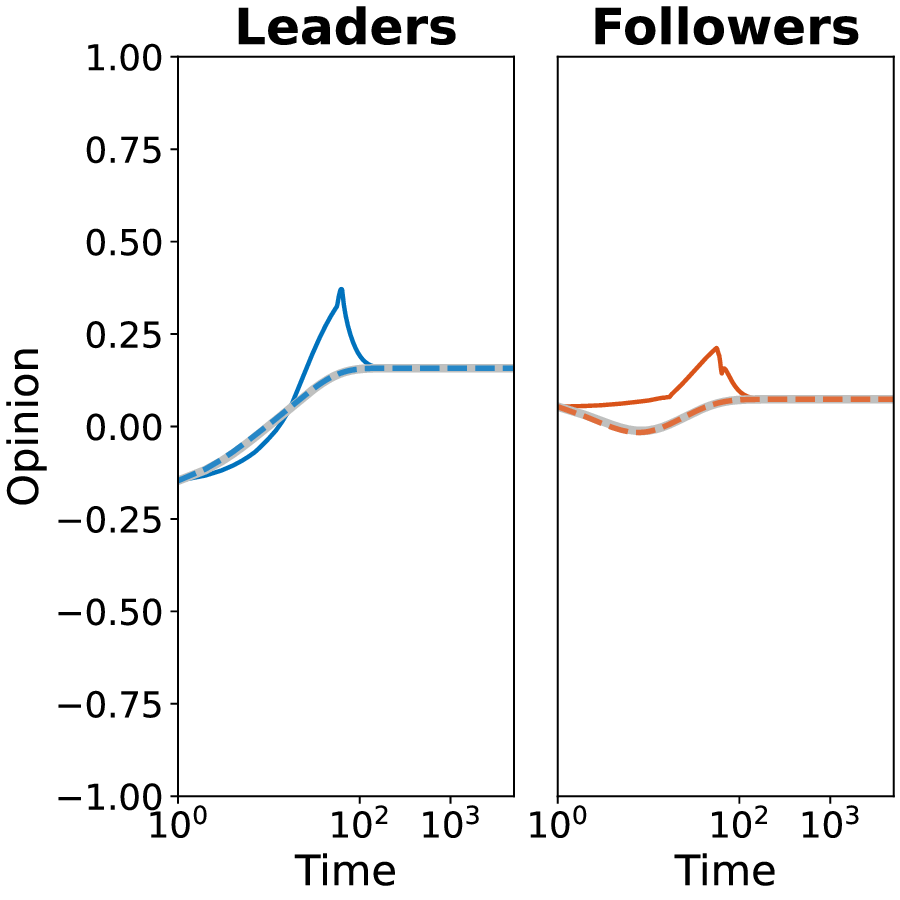}
        \caption{Mean opinion.}\label{fig: comparison-uninformed}
    \end{subfigure}
    \caption{Top row: $N_L=5,\ N_F=6,\ N_U=0$. Bottom row: $N_L=5,\ N_F=6,\ N_U=10.$ The discontinuous grey line in (a), (b) (d) and (e) represents the final mean opinion of leaders and followers. In (c) and (f), the solid blue and red lines represent the numerical results for $r_w=0.5$, the dashed line the numerical results for $\psi\equiv 1$, and the solid grey line the analytical solution.}
    \label{fig:opinions-only-no-uninformed-analytical}
\end{figure}

In Figure \ref{fig:opinions-only-no-uninformed-analytical}, we compare the outcome of the computational experiments for $r_w=0.5$ and $\psi\equiv1$ (therefore $r_w = 2$) with the analytical solution derived in Section \ref{sec:analysis}. Although the rest of the paper uses $r_w=1$ for all numerical computations, we choose $r_w=0.5$ here for a clearer comparison with the case 
$\psi\equiv 1$. The top row of Figure \ref{fig:opinions-only-no-uninformed-analytical} shows the dynamics for $N_U=0$, and the bottom row for $N_U = 10$. The first and second column illustrate the impact of the interaction radius: $r_w =0.5$ (first) vs. $\psi \equiv 1$ (second). We compare the outcome of the computational experiments with the analytical solution in the third column, see Figures \ref{fig: comparison-nouninformed} and \ref{fig: comparison-uninformed}. The computational solution for $\psi\equiv 1$ corresponds to the dashed line, and the analytical solution to the solid grey line. We note that they match perfectly. However, the solution for $r_w=0.5$ (solid blue and red lines) is slightly different from the analytical one, since the analysis did not account for $r_w = 0.5$.

\subsection{Single population swarming with opinion}

We continue by focusing on the single population swarming model \eqref{eq: one-population}. Note that, when excluding the opinion dynamics, \eqref{eq: one-population} reduces to the swarming model of D'Orsogna et al. \cite{d2006self}. This model exhibits rich dynamics, such as mills, flocks, and lumps, depending on the parameters considered. We discuss the dynamics for the different parameter regimes in Appendix \ref{app: swarming model}: see Table \ref{tab: parameters flocking} for parameter values, and Figure \ref{fig: class swarming} for dynamics without opinions, as well as Figure \ref{fig: swarming with op 1pop} for the dynamics in case of the full model.

\noindent In this section, we focus on parameter regimes II and VIII in Table \ref{tab: parameters flocking}, as they lead to the formation of flocks.  Lafoux et al. \cite{lafoux2023illuminance} discussed that the natural behaviour of a school of fish in a tank corresponds to regime II. Therefore, we are going to consider only this case for the three population case, but discuss parameter regimes II and VIII in the single species case. 

\noindent Furthermore, we set $\gamma_T=\gamma_B=1$, $r_w=0.5$, $r_\bx=1$, while the values for $\tau_T$ and $\tau_B$ are specified in each example. All results are presented at time $T=100$.

\subsubsection{Discussion of modelling constraints}
\begin{figure}[!ht]
        \centering
        \includegraphics[height=4cm]{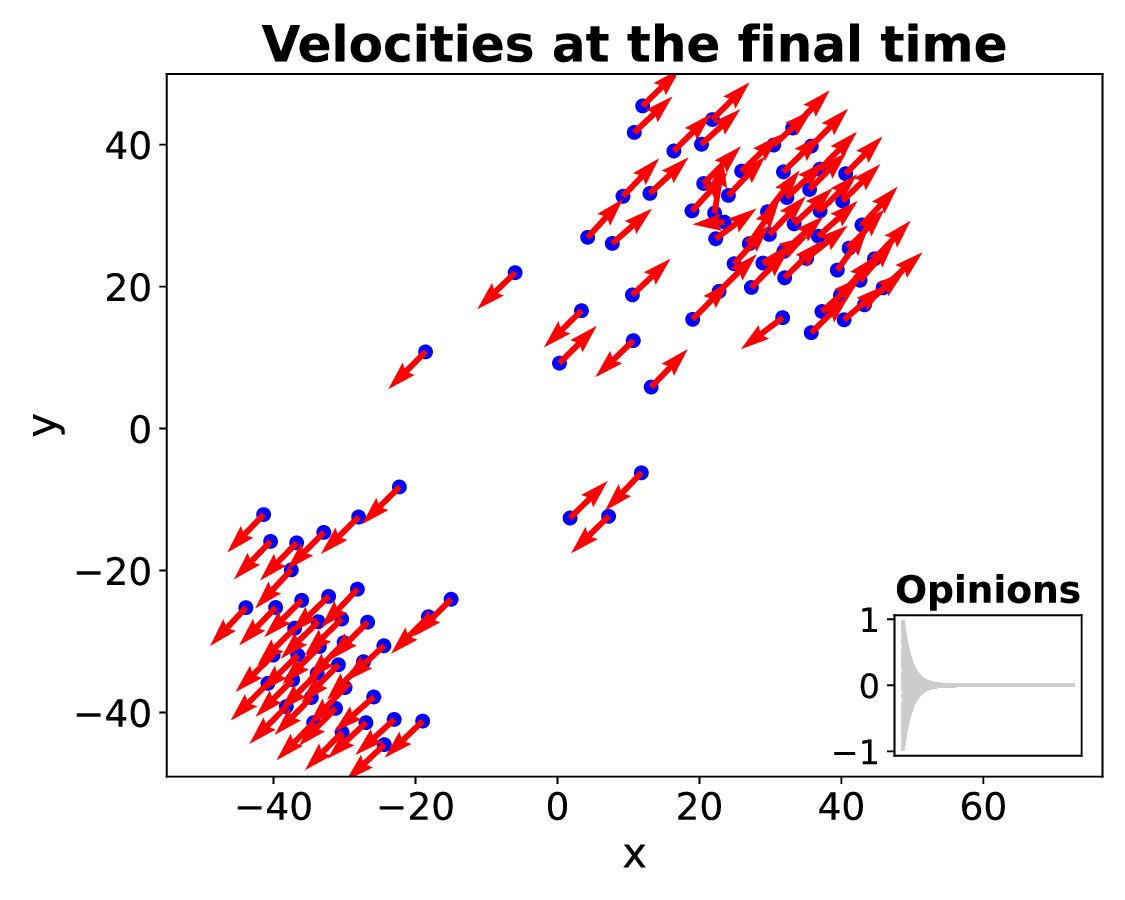}
        \caption{Results of the computational experiments of \eqref{eq: one-population} for $\alpha=1$, $\beta=5$, $C_a=100$, $\ell_a=1.2$, $C_r=350$, $\ell_r=0.8$, $\tau_B=\tau_T=0.1$, $r_\bx=0.5$ and $r_w=1$.}
        \label{fig:one-pop-case-viii-d}
\end{figure}

Before discussing the results, we would like to comment on specific choices of modelling parameters.
For the one population model with opinion alignment \eqref{eq: one-population}, in order to see spatial polarisation, i.e., the case when the swarm splits and goes in opposite directions, we need to choose $r_\bx\ll1$, for a random initial conditions in the square $[-1,1]\times[-1,1]$, and the parameters in the function \eqref{eq: morse potential} appropriately. 
For instance, in Figure \ref{fig:one-pop-case-viii-d} we have the solution of \eqref{eq: one-population} for the parameters specified in the figure caption.
Here, we observe consensus in opinion, but polarisation in velocity.  Opinion promotes the velocity alignment through the last term in equation \eqref{eq: velocity1}. On the other hand, the interaction potential $\mathcal{U}$ produces a clustering due to the high attraction coefficient ($C_a=100$) and short attraction distance ($\ell_a=1.2$), while there is a strong repulsion ($C_r=350$) among individuals at short distance ($\ell_r=0.8$). Once the clusters are formed, they keep moving towards the preferred velocity (either $\bv_T=(1,1)$ or $\bv_B=(-1,-1)$) influenced by the opinion until a point where the attraction component of the potential $\mathcal{U}$ has no effect.

Even choosing the same strong conviction for both populations, $\tau_T=\tau_B=1$, is not sufficient to observe the splitting of the swarm. This is because  this model is highly based on alignment and attraction-repulsion forces. 

To account for the fact that two individuals might be forced to interact in space despite their opinion preferences being far from each other, we choose $r_w<r_\bx$, that is, for two individuals to interact, their opinion must be relatively closer than their position. In general, we might consider 
$$\frac{r_w}{|w_T-w_B|}<\frac{r_\bx}{|\bv_T-\bv_B|}\ .$$

In the three-population model, one could introduce an opinion interaction radius $r_{w}$ that depends on the interacting populations, denoted as $r_w^{pq}$. For example, leader–follower or leader–uninformed interactions might feature a larger $r_{w}$, reflecting the stronger influence leaders exert, even when follower or uninformed opinions are relatively distant. Conversely, follower–uninformed interactions could have a smaller value of $r_{w}$. However, for simplicity, we assume the same value of $r_{w}$ across all interactions.\\

\begin{figure}[!ht]
    \centering
    \begin{subfigure}{0.32\linewidth}
        \centering
        \includegraphics[height=4cm]{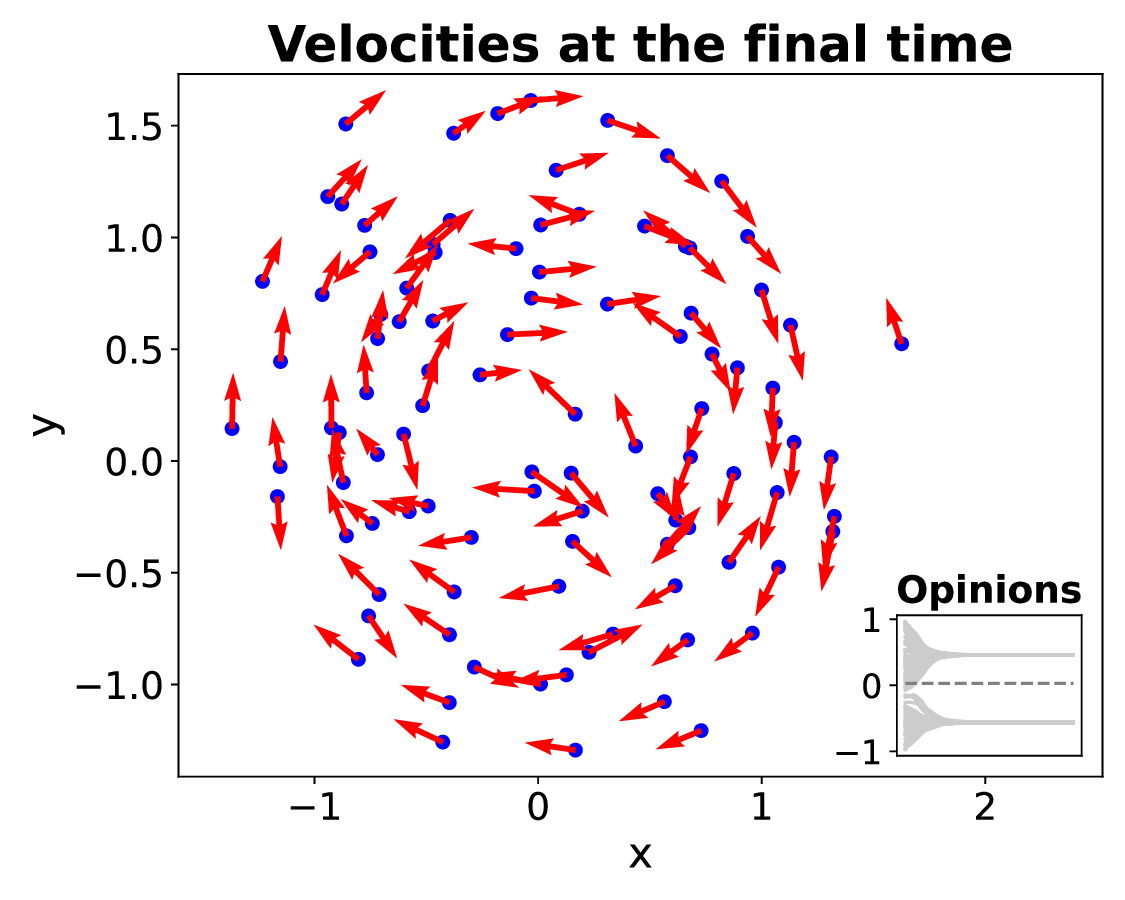}
        \caption{$\tau_T=\tau_B=0$}
        \label{fig:one-pop-case-ii-a}
    \end{subfigure}
    \begin{subfigure}{0.32\linewidth}
        \centering
        \includegraphics[height=4cm]{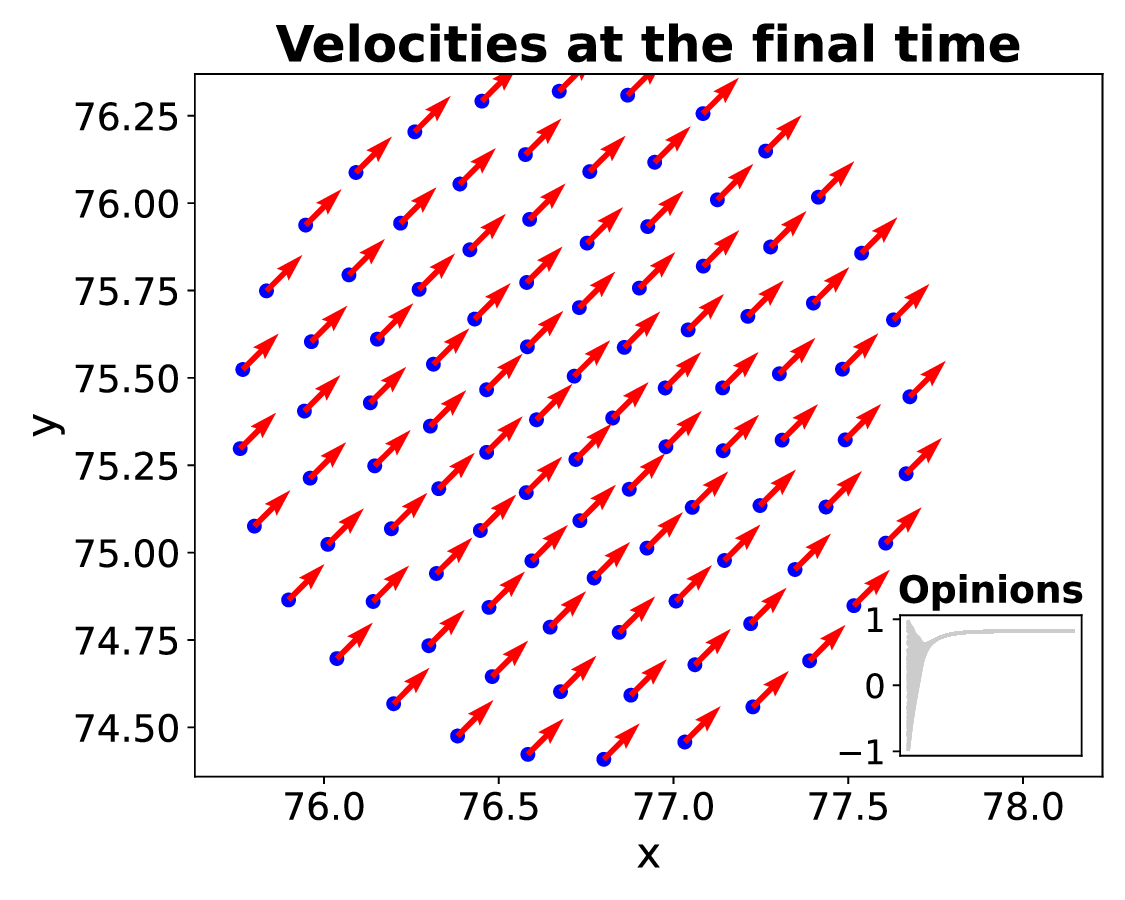}
        \caption{$\tau_T=0.1$ and $\tau_B=0.01$}
        \label{fig:one-pop-case-ii-b}
    \end{subfigure}
    \begin{subfigure}[b]{0.32\linewidth}
        \centering
        \includegraphics[height=4cm]{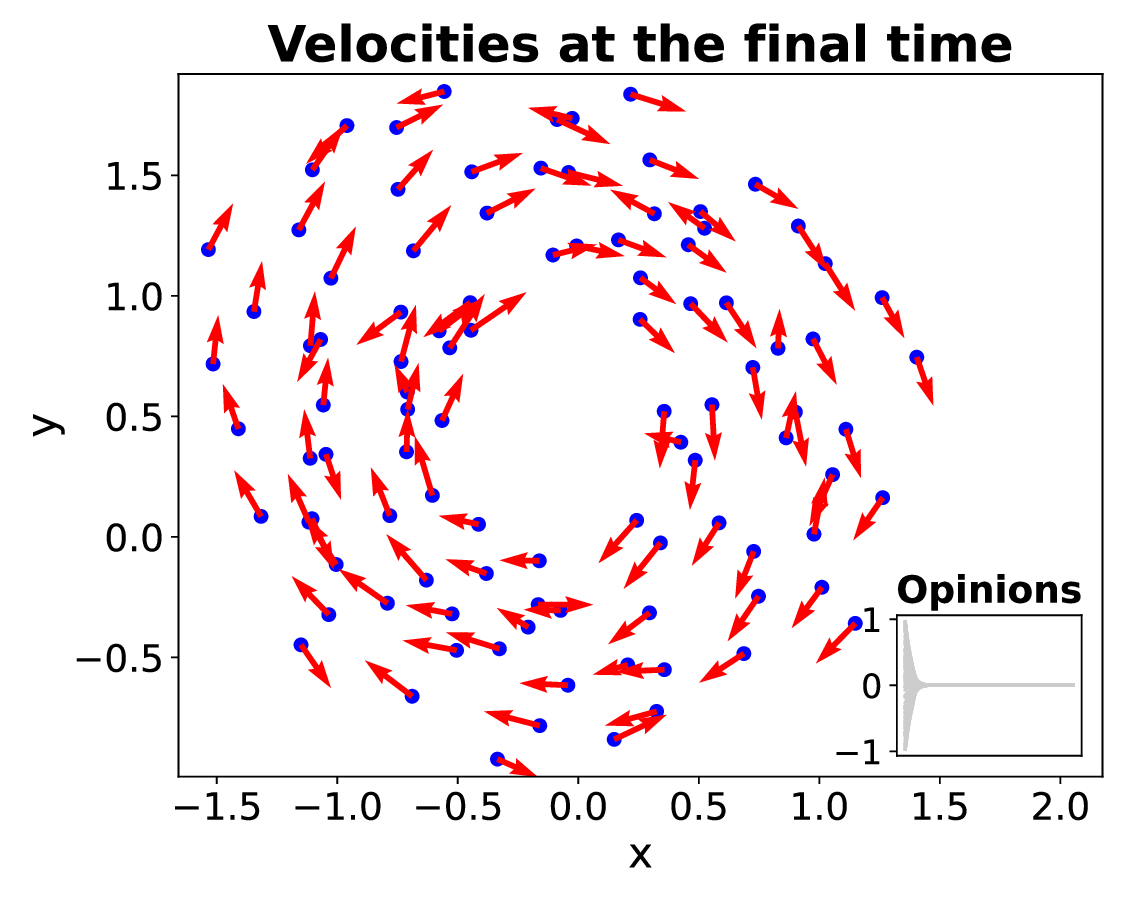}
        \caption{$\tau_T=\tau_B=0.1$}
        \label{fig:one-pop-case-ii-c}
    \end{subfigure}
    \caption{
        Results of the computational experiments for parameter regime II ($\alpha=1,\ \beta=0.5,\ C_a=50,\ \ell_a=1,\ C_r=60, \ \ell_r=0.5$) with $N=100$, {$r_\bx=1$ and $r_w=0.5$}.
    }
    \label{fig:one-pop-case-ii}
\end{figure}

\begin{figure}[!ht]
    \centering    
\begin{subfigure}{0.32\linewidth}
        \centering
        \includegraphics[height=4cm]{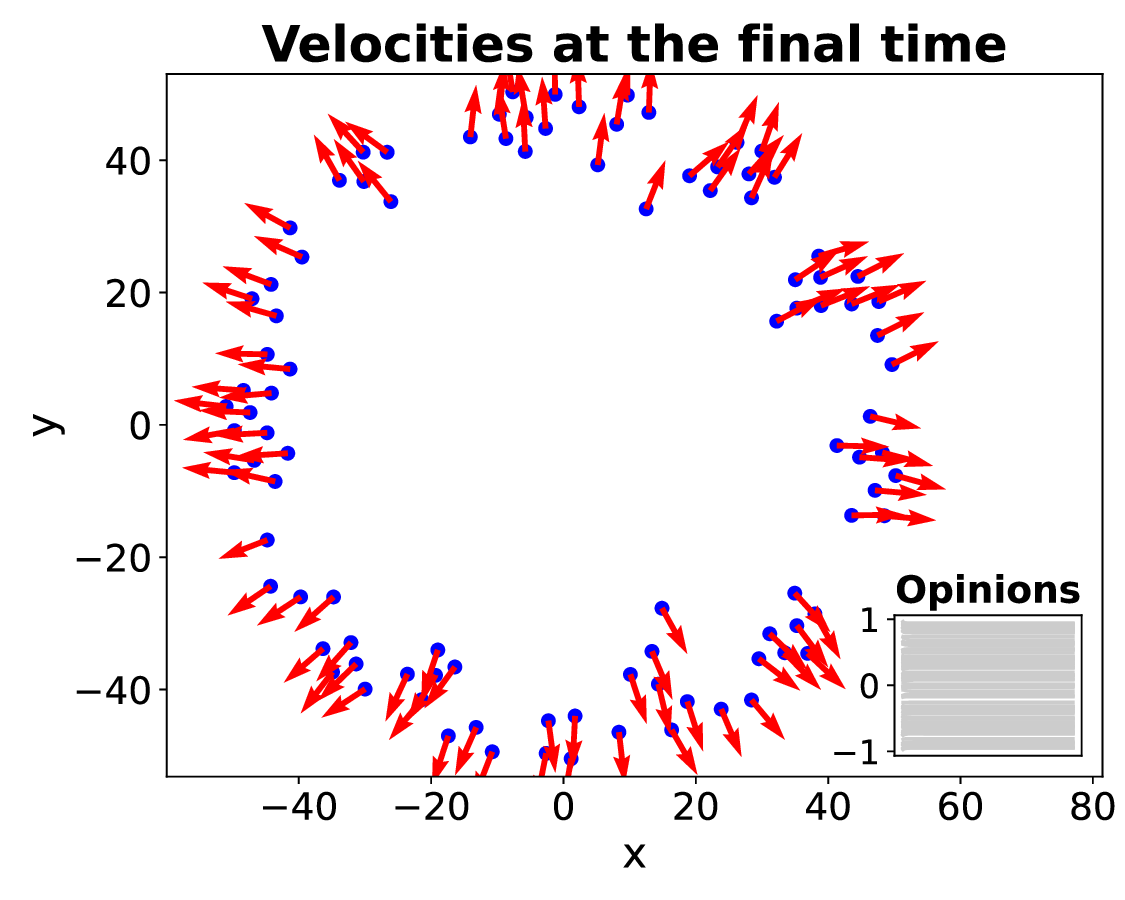}
        \caption{$\tau_T=\tau_B=0$}
        \label{fig:one-pop-case-viii-a}
    \end{subfigure}
\begin{subfigure}{0.32\linewidth}
        \centering
        \includegraphics[height=4cm]{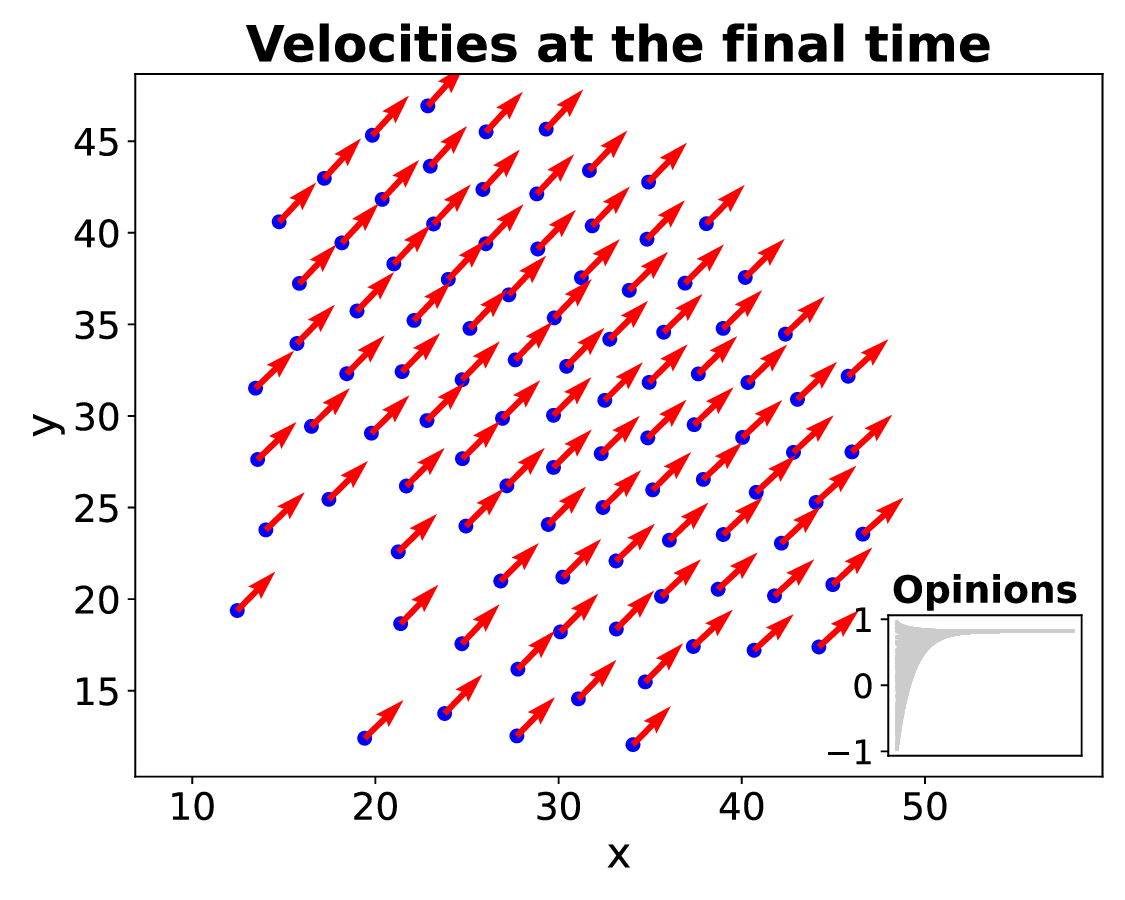}
        \caption{$\tau_T=0.1$ and $\tau_B=0.01$}
        \label{fig:one-pop-case-viii-b}
    \end{subfigure}
\begin{subfigure}{0.32\linewidth}
        \centering
        \includegraphics[height=4cm]{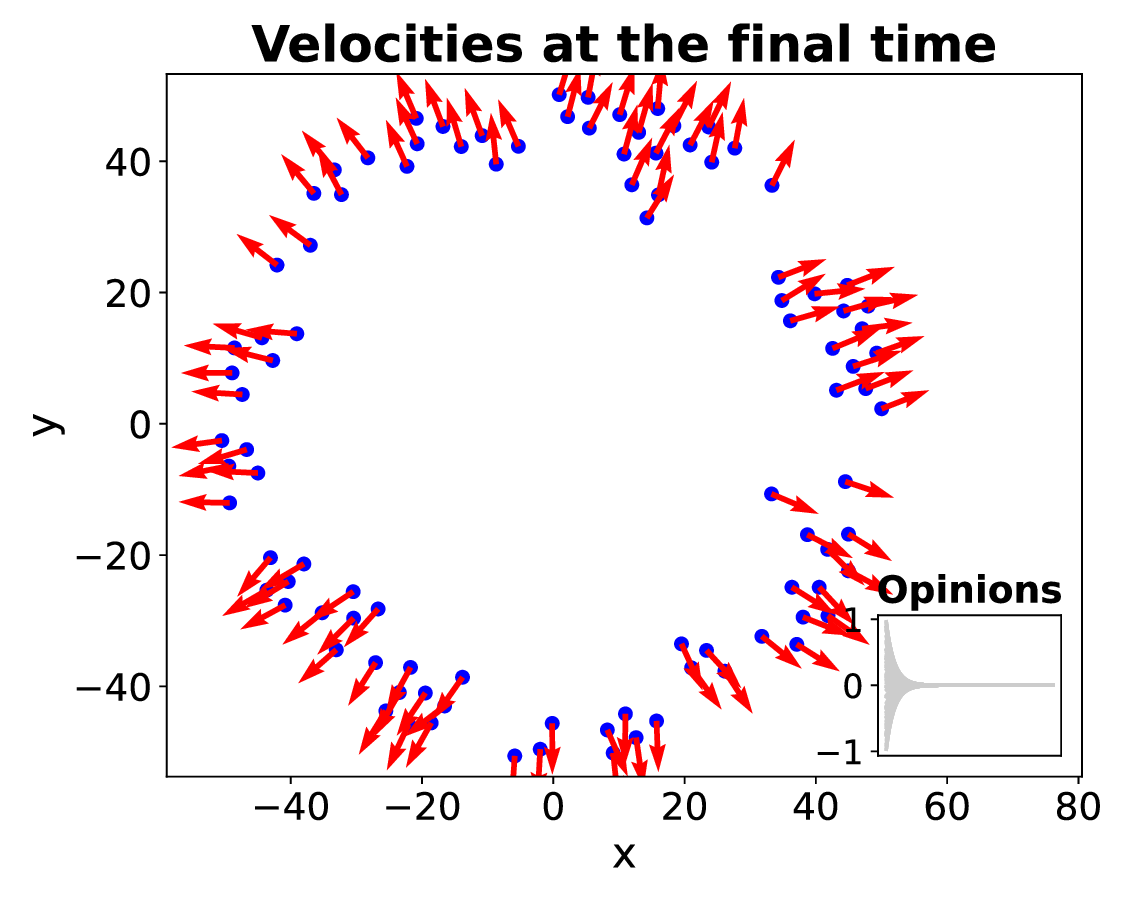}
        \caption{$\tau_T=\tau_B=0.1$}
        \label{fig:one-pop-case-viii-c}
    \end{subfigure}
    \caption{
        Results of the computational experiments for parameter regime VIII ($\alpha=1,\ \beta=5,\ C_a=100,\ \ell_a=1.2,\ C_r=350,\ \ell_r=0.8$) with $N=100$, $r_\bx=1$, $r_{w}=0.5$. 
    }
    \label{fig:one-pop-case-viii}
\end{figure}

Figure \ref{fig:one-pop-case-ii} shows the outcomes of the computational experiments in parameter regime II for different choices of $\tau_T$ and $\tau_B$, while Figure \ref{fig:one-pop-case-viii} presents the same scenarios for parameter regime VIII. In all cases, the embedded figures show the evolution of opinion. In Figures \ref{fig:one-pop-case-ii-a} and \ref{fig:one-pop-case-viii-a}, individuals have no preference ($\tau_T = \tau_B = 0$) -- we observe mill formation in space and cluster formation in opinion in Figure \ref{fig:one-pop-case-ii-a}, while we see splitting dynamics and no opinion alignment in Figure \ref{fig:one-pop-case-viii-a}. The latter is caused by the spatial dependence in opinion dynamics -- since the swarm spreads out, individuals are too far apart to interact in opinion (preventing consensus formation).

In Figures \ref{fig:one-pop-case-ii-b} and \ref{fig:one-pop-case-viii-b}, some individuals exhibit a stronger preference towards the top target, and others show a weaker preference for the bottom target ($\tau_T = 0.1$, $\tau_B = 0.01$). Let us recall that we are still in the case of one population. In this case, the alignment in opinion, showed in the embedded figures where we observe consensus towards $1$,  leads to an alignment of velocities in the direction of $\mathbf{v}_T = (1,1)$. We observe that the cluster formation in Figure \ref{fig:one-pop-case-ii-b} is more compact, therefore the opinion converges faster than in Figure \ref{fig:one-pop-case-viii-b}, where the cluster is more elongated. The size of the cluster is determined by the potential $\mathcal{U}$. 

Finally, in Figures \ref{fig:one-pop-case-ii-c} and \ref{fig:one-pop-case-viii-c}, all individuals share the same preference ($\tau_T = \tau_B = 0.1$). This case is similar to $\tau_T=\tau_B=0$, only  consensus in opinion is achieved.  We notice that the mill in Figure \ref{fig:one-pop-case-ii-a}  has a more egg-like shape due to the polarisation in opinion. 

We recall the conditions of the cohesive motion steady state derived in Subsection \ref{sec: opinion-swarming analytic}. For the case in Figure \ref{fig:one-pop-case-ii-b}, we know that $\alpha=1$, $\beta=0.5$, $r_\bx=1$ and $r_w=0.5$ and the opinion steady state is $w^*\approx 1$. Since $w_T=1$ and $w_B=-1$, we have $\psi(|w_T-w^*|)=1$ for $r_w=0.5$ and zero otherwise. Therefore, as explained in Remark \ref{rem: steady state velo},  the velocity steady state \eqref{eq: one velocity steady state} becomes 
\[
    \bv^*=\frac{\bv_T}{\beta|\bv^*|^2}\ ,
\]
which is a vector that points in the positive direction in the first quadrant, according to the numerical results  in Figure \ref{fig:one-pop-case-ii-b}. A similar analysis can be done if $w^*\approx-1$ and therefore individuals move in the negative direction. \\

\subsection{Three population swarming with opinion}\label{sec: 3 pop swarming with opinions}

We conclude by investigating the behaviour of the full three population swarming opinion model \eqref{eq: opinion_sawarm 3pop}. In the following we consider parameter regimes II, that is  $r_\bx=1$, $r_w=0.5$, $\gamma_T^p=\gamma_B^p=1$ and all $u_{pq}=k_{pq}=1$ unless specified otherwise. The uninformed's opinions are initialised to $0$, while the rest are sampled uniformly from $[-1,1]$. All plots show the results at time $T=100$.\\

\begin{figure}[!ht]
    \centering
    \begin{subfigure}[b]{0.32\linewidth}
        \centering
        \includegraphics[height=4cm]{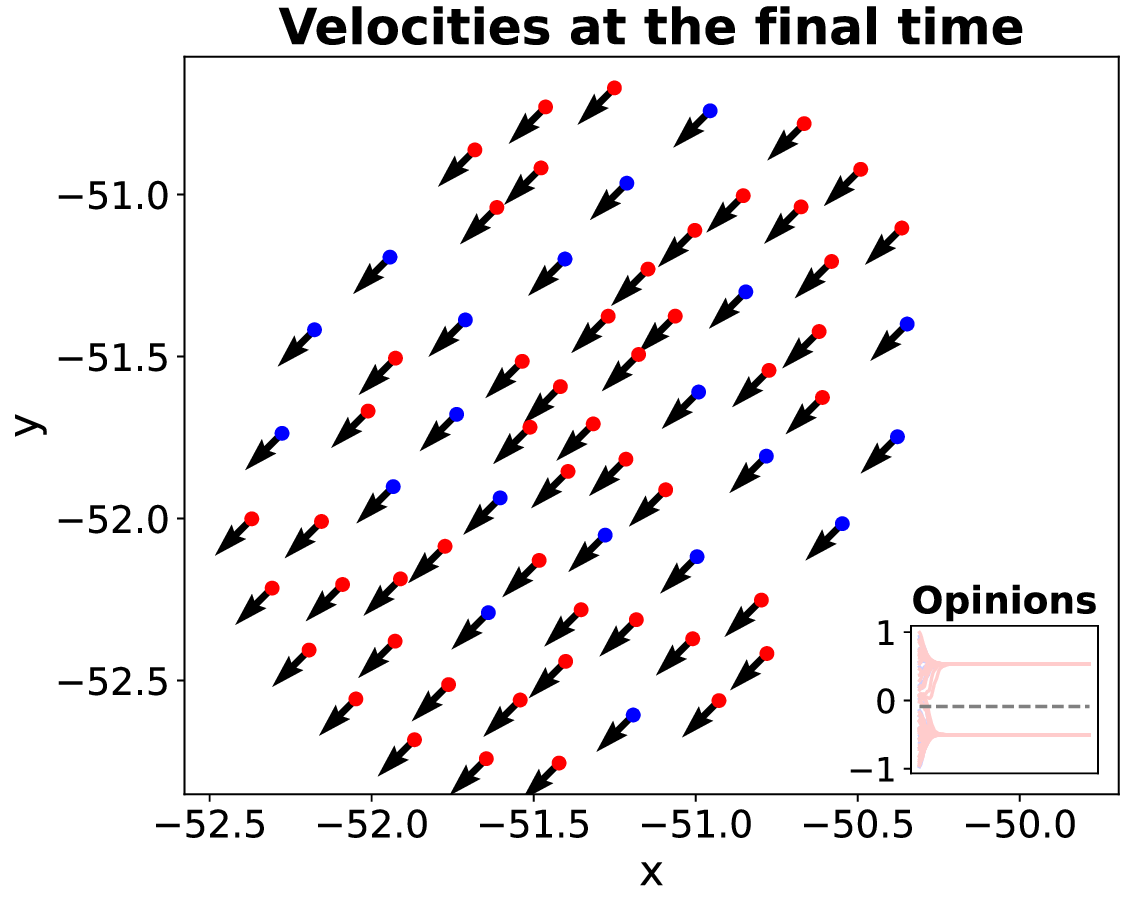}
        \caption{}
        \label{fig:three-pop-1a}
    \end{subfigure}
    \begin{subfigure}[b]{0.32\linewidth}
        \centering
        \includegraphics[height=4cm]{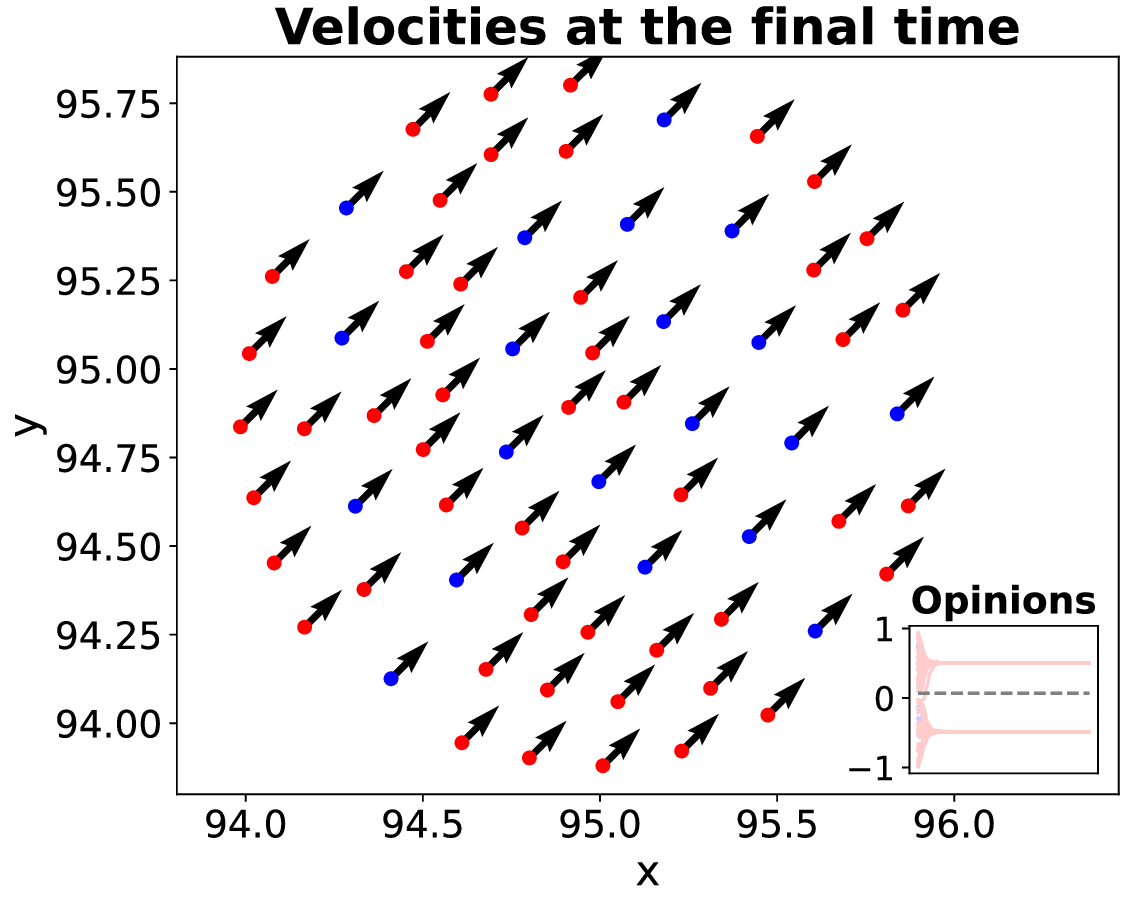}
        \caption{}
        \label{fig:three-pop-1b}
    \end{subfigure}
    \begin{subfigure}{0.32\linewidth}
            \centering
            \includegraphics[height=2cm]{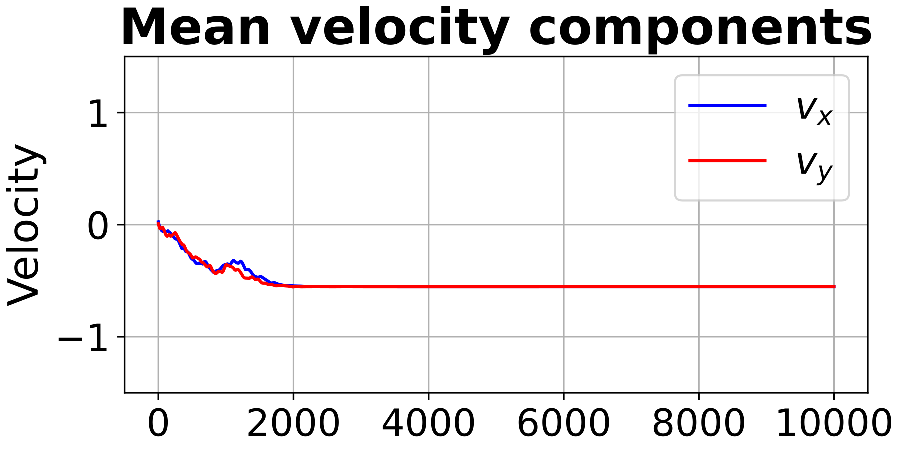}
            \includegraphics[height=2cm]{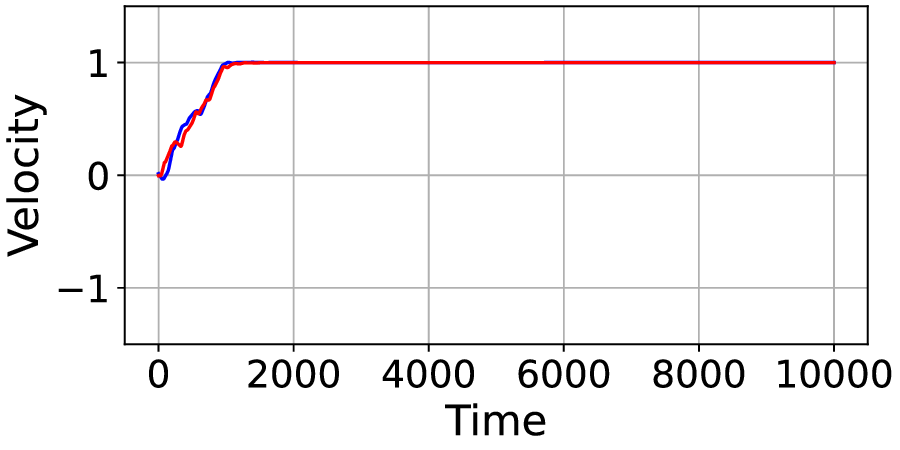}
            \caption{}
            \label{fig:third-top}
        \end{subfigure}
    \caption{ Velocity alignment: (a) Since the average initial opinion is negative, the flock aligns in the negative direction. (b) Since the average opinion is positive, the flock aligns in the positive direction. (c) Shows the mean velocity components ($v_x,\ v_y$) for the result in (a) (top), and in (b) (bottom).
        Here $N_L=20$ (in blue), $N_F=50$ (in red), and $N_U=0$. We consider no preferences $\tau_T=\tau_B=0$, and parameters $k_{LL}=k_{FF}=k_{LF}=k_{FL}=1$. 
    }
    \label{fig:three-pop-1}
\end{figure}

\begin{figure}[!ht]
    \centering
    \begin{subfigure}[t]{0.4\linewidth}
        \centering
        \includegraphics[height=4cm]{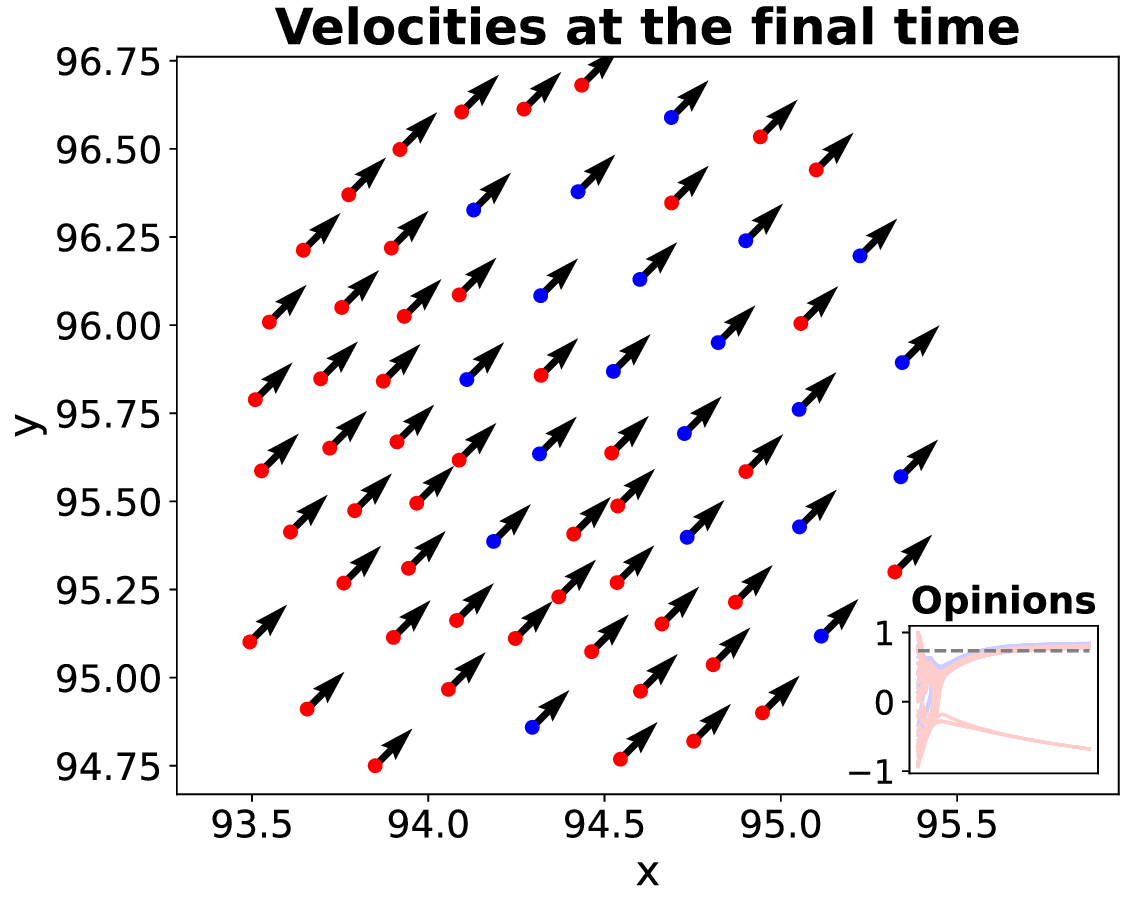}
        \caption{Follower-leaders only.}
        \label{fig:three-pop-2}
    \end{subfigure}
    \hspace{1em}
    \begin{subfigure}[t]{0.4\linewidth}
        \centering
        \includegraphics[height=4cm]{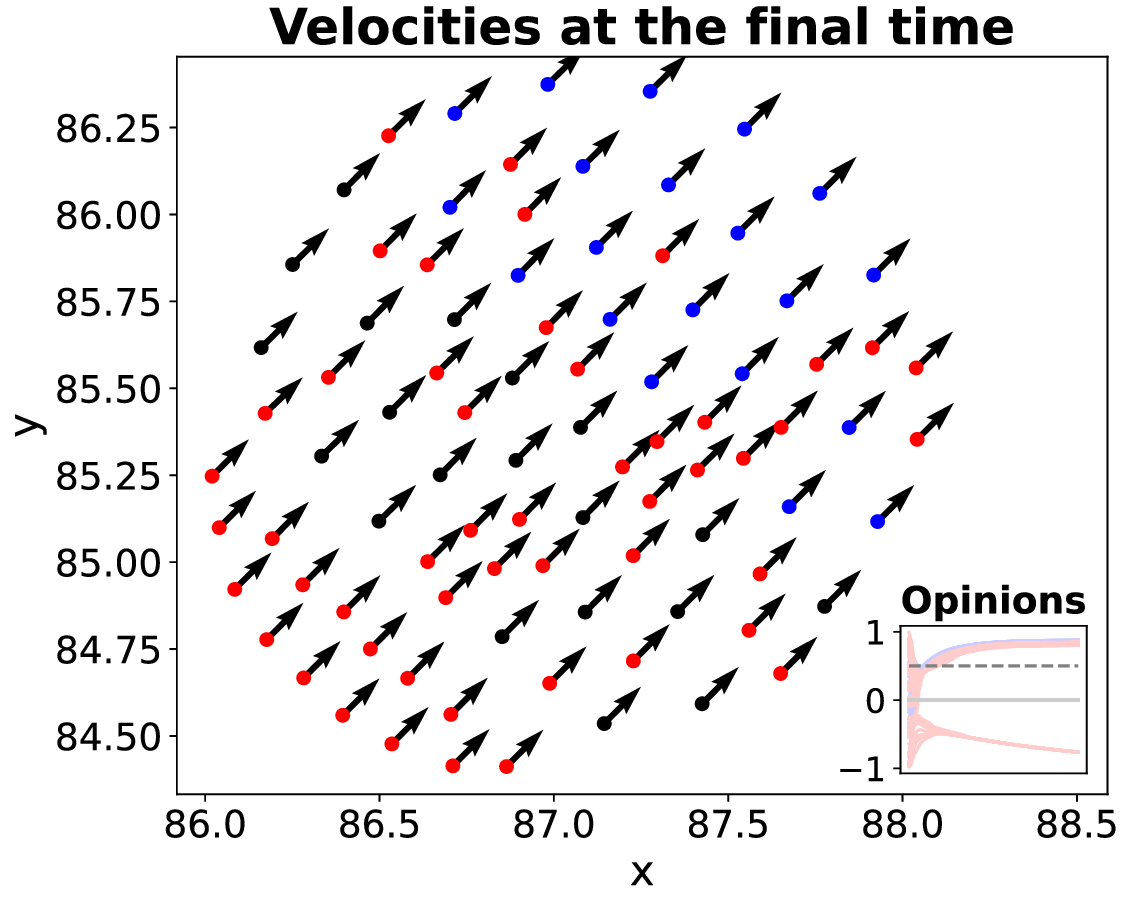}
        \caption{Follower, leaders and uninformed.}
        \label{fig:three-pop-3}
    \end{subfigure}
    \caption{
        Leaders, in blue, have a strong preference for the top target, $\tau_T=0.1$, while followers, in red, have a weak preference for the bottom target, $\tau_B=0.01$. Uninformed individuals, in black, do not have any preference. Here $N_L=20$, $N_F=50$ and $N_U=20$. The parameters are set to $k_{LL}=k_{FF}=k_{LF}=k_{FL}=1$ and in (b) $k_{Up}=k_{pU}=0$ for all $p$. 
    }
    \label{fig:three-pop-2-3}
\end{figure}

\begin{figure}[!ht]
    \centering
    \begin{subfigure}[t]{0.32\linewidth}
        \centering
        \includegraphics[height=4cm]{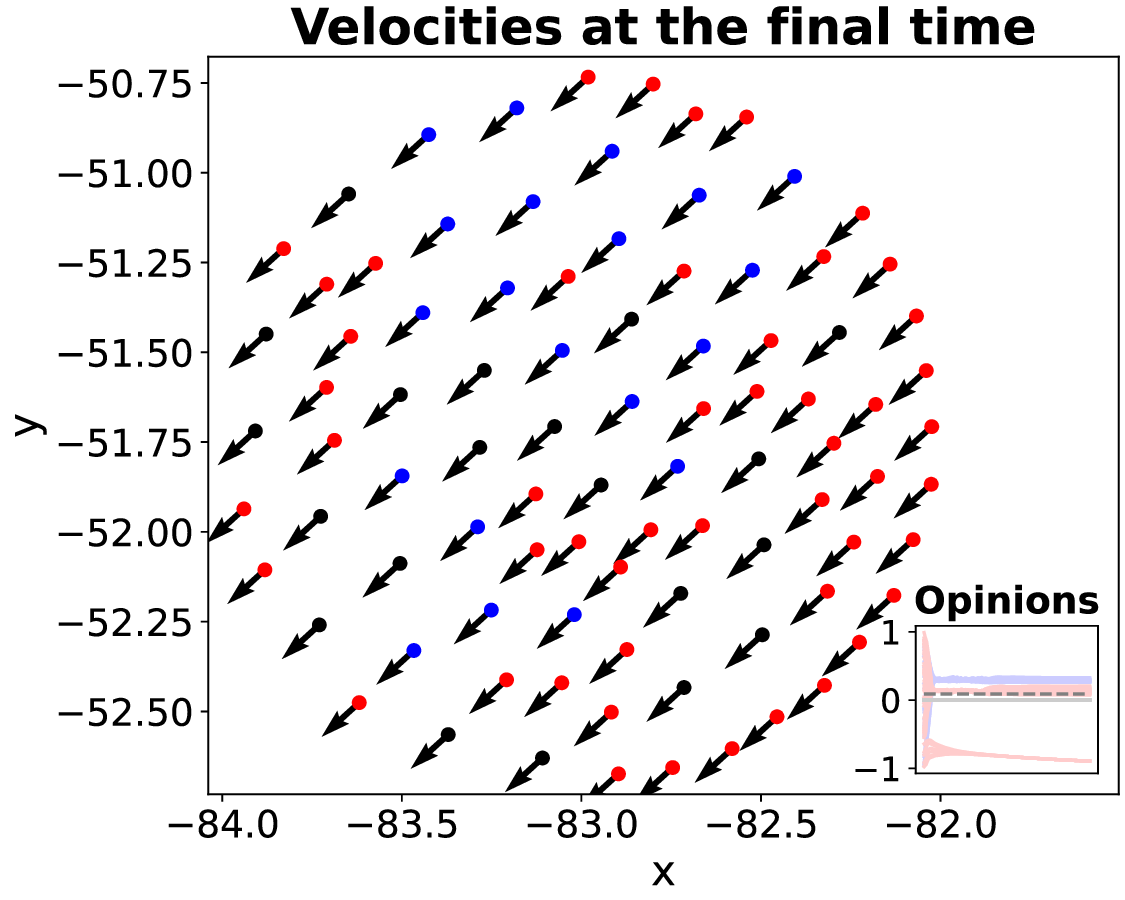}
        \caption{Flock moves towards the bottom.}
        \label{fig:three-pop-4a}
    \end{subfigure}
    \begin{subfigure}[t]{0.32\linewidth}
        \centering
        \includegraphics[height=4cm]{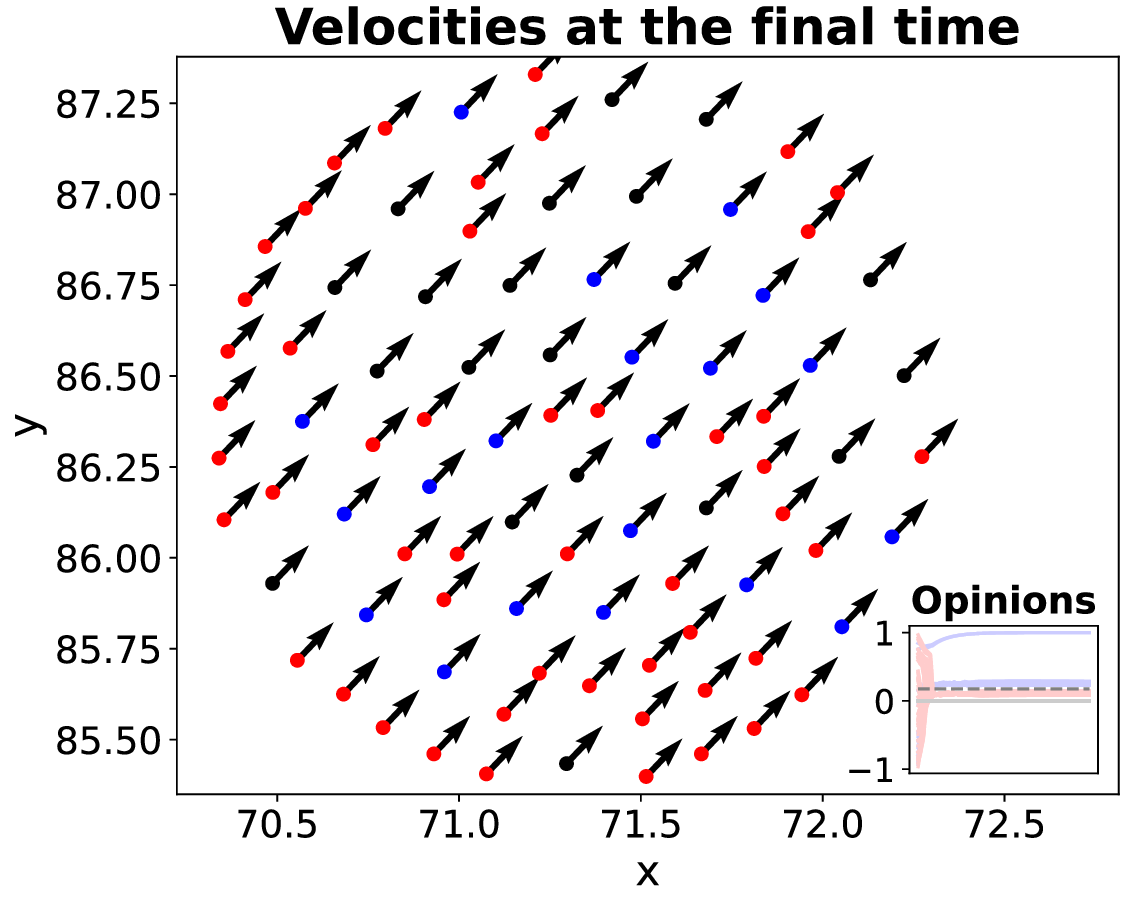}
        \caption{Flock moves toward the top.}
        \label{fig:three-pop-4b}
    \end{subfigure}
    \begin{subfigure}[t]{0.32\linewidth}
        \centering
        \includegraphics[height=4cm]{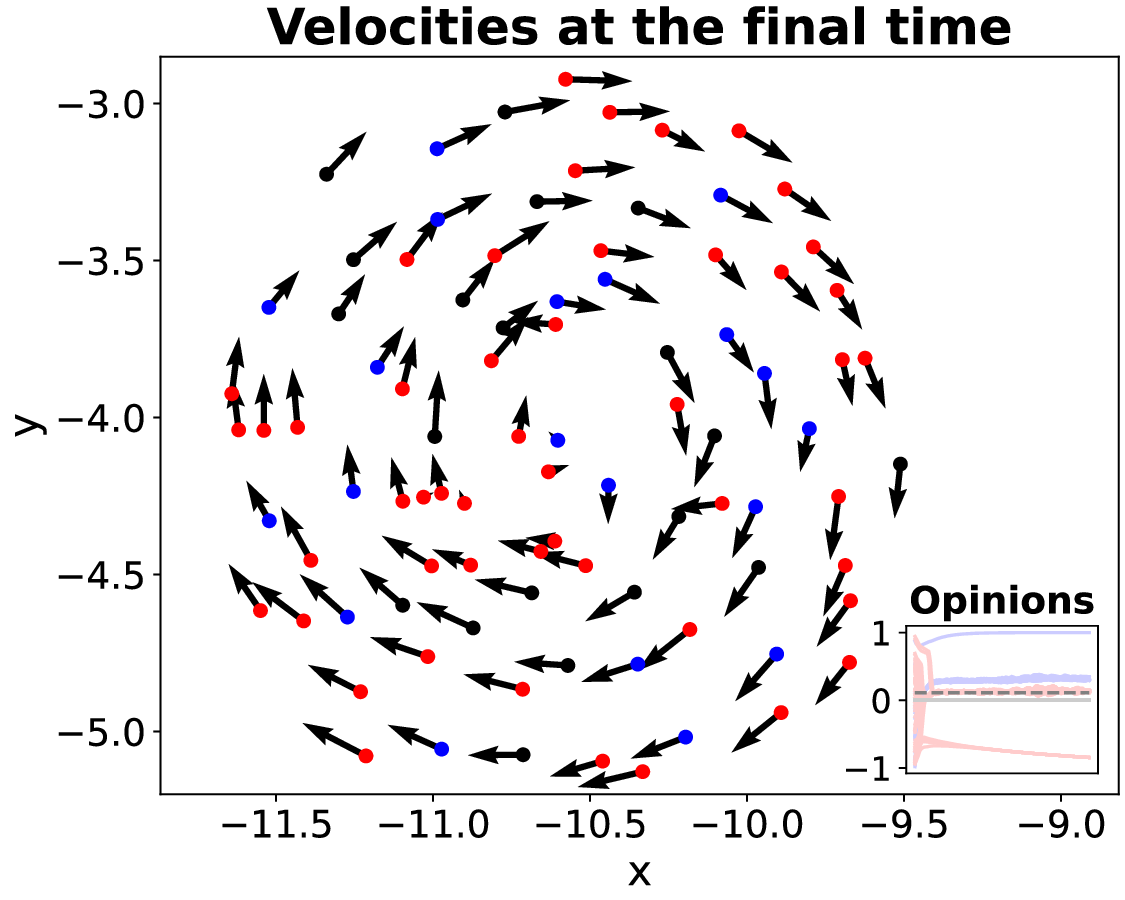}
        \caption{Mill dynamics.}
        \label{fig:three-pop-4c}
    \end{subfigure}
    \caption{
        Leaders, in blue, have a preference given by $\tau_T=0.1$, while followers, in red, have a preference given by $\tau_B=0.01$, and  uninformed in black have no preference. Here $N_L=20$, $N_F=50$ and   $N_U=20$ with $k_{UU}=k_{UF}=k_{UL}=k_{LU}=0,\ k_{pq}=1$ otherwise. 
    }
    \label{fig:three-pop-4}
\end{figure}

In this section, we illustrate the dynamics in the following three cases:
\begin{enumerate}
    \item Dynamics of followers and leaders only (no uninformed) for $\tau_B = \tau_F = 0$ (see Figure \ref{fig:three-pop-1}).
    \item Impact of the presence of  uninformed and preferential direction on the dynamics. The uninformed do not interact in opinion with the rest (see Figure \ref{fig:three-pop-2-3}).
    \item  Impact of followers interacting in opinion with uninformed on the dynamics (see Figure \ref{fig:three-pop-4}).
\end{enumerate}
Followers correspond to red dots, leaders to blue dots, and uninformed to black dots.\\

Case 1: If followers and leaders have no preference for either direction, the final alignment of the flow is determined by the initial opinion distribution, see Figure \ref{fig:three-pop-1}. If the average initial opinion is negative (inset Figure \ref{fig:three-pop-1a}), the mean velocity of the individuals is $\bv_B=(-1,-1)$, while if the mean average opinion is positive, the swarm moves in $\bv_T=(1,1)$ (Figure \ref{fig:three-pop-1b}). Figure \ref{fig:third-top} shows the mean velocity components for both cases -- we observe very fast alignment.\\

Case 2: Next we investigate the impact of a stronger preference on the outcome, that is, $\tau_T=0.1$ and $\tau_B=0.01$. In Figure \ref{fig:three-pop-2} we observe the same follower-leader dynamics as in the opinion only Example 1 in Section \ref{sec: numerics opinion only}, where the flock moves towards the top driven by the preference of the leaders. We also observe that some individuals (red lines in the low part of the opinion plot) are left behind in opinion due to the choice of $r_\bx$ and $r_{w}$. These individuals correspond to the three red dots in the bottom left corner of the velocity plot in Figure \ref{fig:three-pop-2}.\\
If uninformed individuals are included, but they do not interact in opinion, then they only align in velocity to the rest of the flock, see Figure \ref{fig:three-pop-3}. The introduction of this population does not affect the overall opinion dynamics and therefore the final velocity is again dominated by the leaders' preference.\\

Case 3: If uninformed interact with followers (but not with leaders), that is,  $k_{UL}=k_{UF}=k_{UU}=0$, and $k_{FU}=1$,  three cases arise depending on the initial conditions: (i) Prevalence of the followers' opinion (Figure \ref{fig:three-pop-4a}), (ii) Prevalence of the leaders' opinion (Figure \ref{fig:three-pop-4b}), and (iii) Mill dynamics (Figure \ref{fig:three-pop-4b}).

 To better understand the dynamics in Figure \ref{fig:three-pop-4}, we compute the polarisation and momentum of the swarm defined as in \cite{chuang2007state},
\begin{equation}\label{eq: pol and mom}
    P=\frac{\left| \sum\limits_{i=1}^N\bv_i\right|}{\sum\limits_{i=1}^N|\bv_i|}\ ,\quad M=\frac{\left| \sum\limits_{i=1}^N\mathbf{r}_i\times\bv_i\right|}{\sum\limits_{i=1}^N|\mathbf{r}_i||\bv_i|}\ ,
\end{equation}
where  $\mathbf{r}_i=\bx_i-\bx_{CM}$ is the distance between the $i$-th individual and the centre of mass of the flock.
 
\begin{figure}[!ht]
    \centering
    \begin{subfigure}[t]{0.32\linewidth}
        \centering
        \includegraphics[height=4cm]{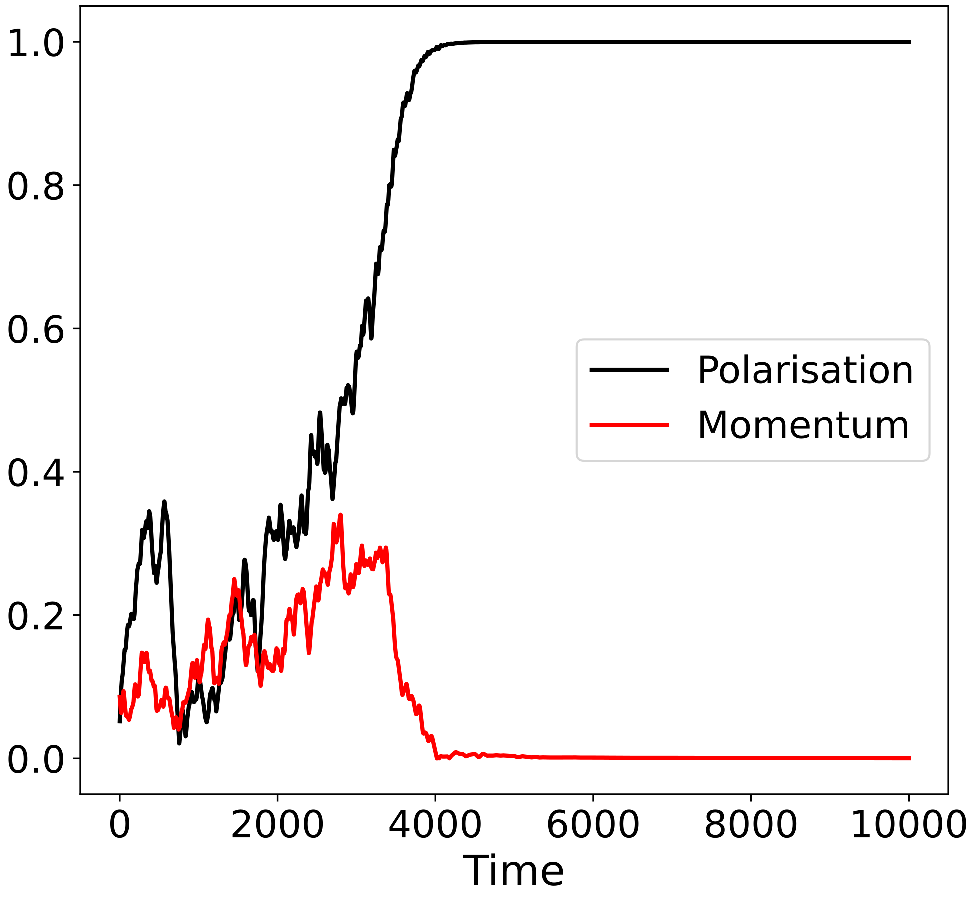}
        \caption{Flock moves towards the bottom.}
        \label{fig:three-pop-4a-quantities}
    \end{subfigure}
    % \begin{subfigure}[t]{0.32\linewidth}
    %     \centering
    %     \includegraphics[height=4cm]{case-4b_quantities.svg}
    %     \caption{Flock to the top.}
    %     \label{fig:three-pop-4b-quantities}
    % \end{subfigure}
    \begin{subfigure}[t]{0.32\linewidth}
        \centering
        \includegraphics[height=4cm]{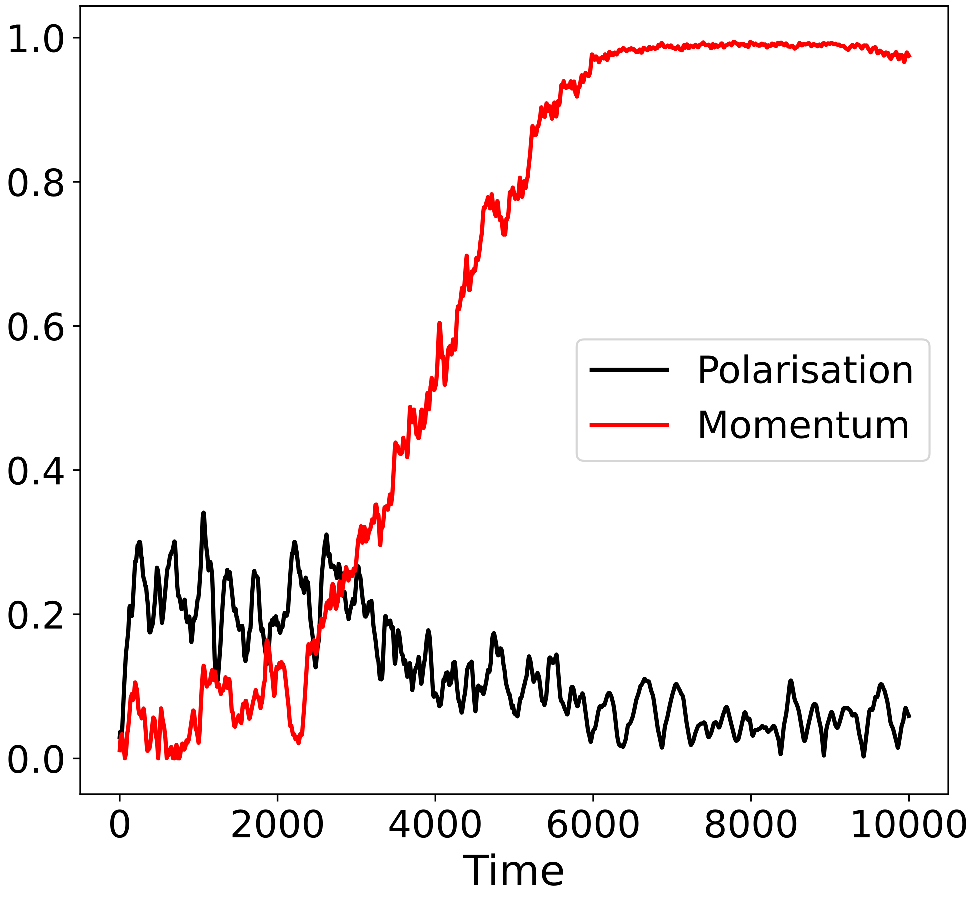}
        \caption{Mill dynamics.}
        \label{fig:three-pop-4c-quantities}
    \end{subfigure}
    \caption{
        Polarisation and momentum over time corresponding to Figure \ref{fig:three-pop-4}.     }
    \label{fig:three-pop-4-quantities}
\end{figure}

Quantities in  \eqref{eq: pol and mom} are computed in Figure \ref{fig:three-pop-4-quantities}, to confirm that the system does not settle into meta-stable configurations. In Figure \ref{fig:three-pop-4a-quantities} we observe that the polarisation reaches $1$ after some time and remains constant, while the momentum approaches $0$, corresponding to Figures \ref{fig:three-pop-4a} and \ref{fig:three-pop-4b}. In contrast, for Figure \ref{fig:three-pop-4c-quantities}, the momentum increases toward $1$, whereas the polarisation oscillates around $0$.\\
 
\begin{figure}[!ht]
    \centering
    \begin{subfigure}[t]{0.4\linewidth}
        \centering
         \includegraphics[height=4.5cm]{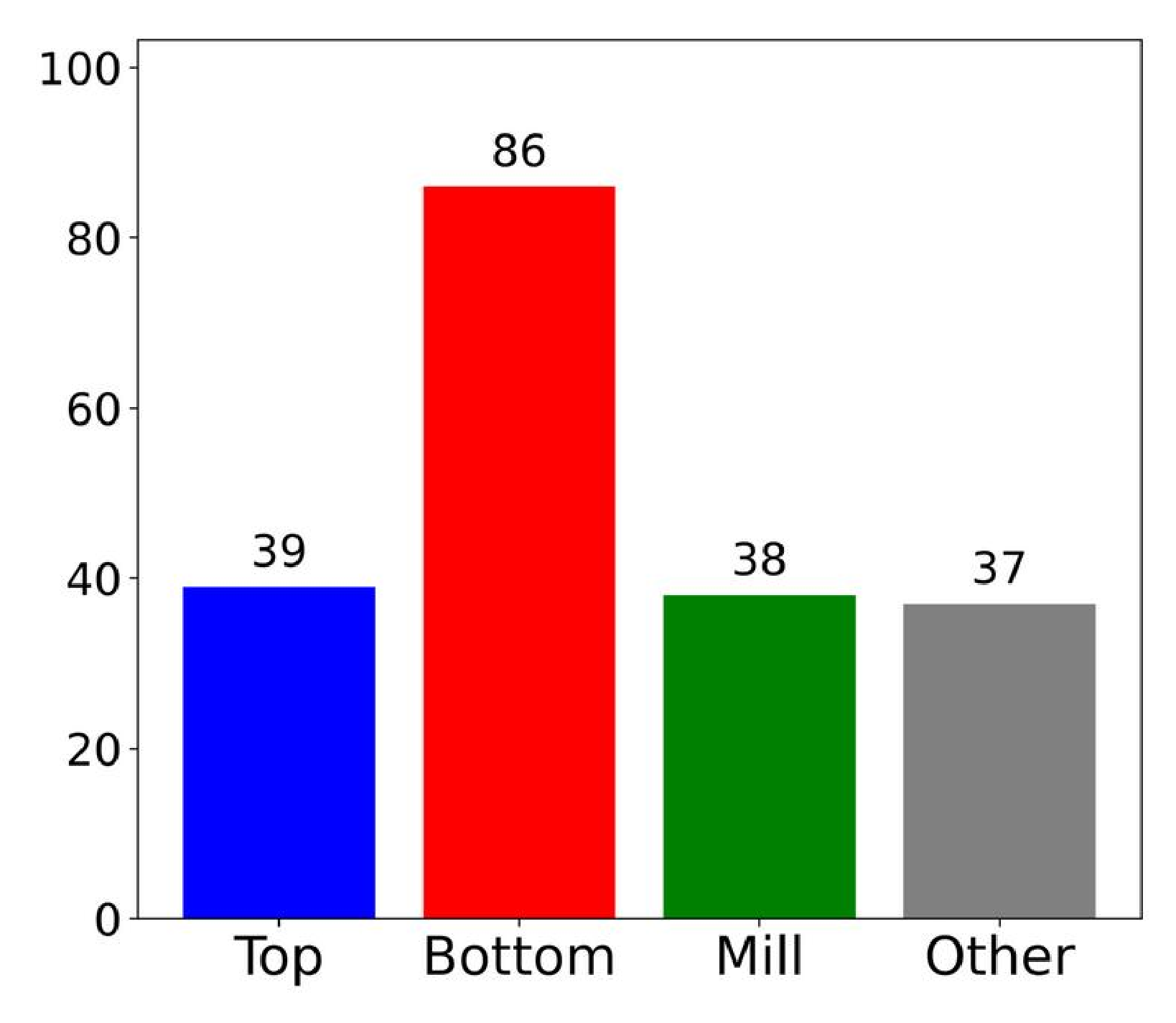}       
        \caption{}
        \label{fig: percentages-a}
    \end{subfigure}
    \hspace{1em}
    \begin{subfigure}[t]{0.4\linewidth}
        \centering
        \includegraphics[height=4.5cm]{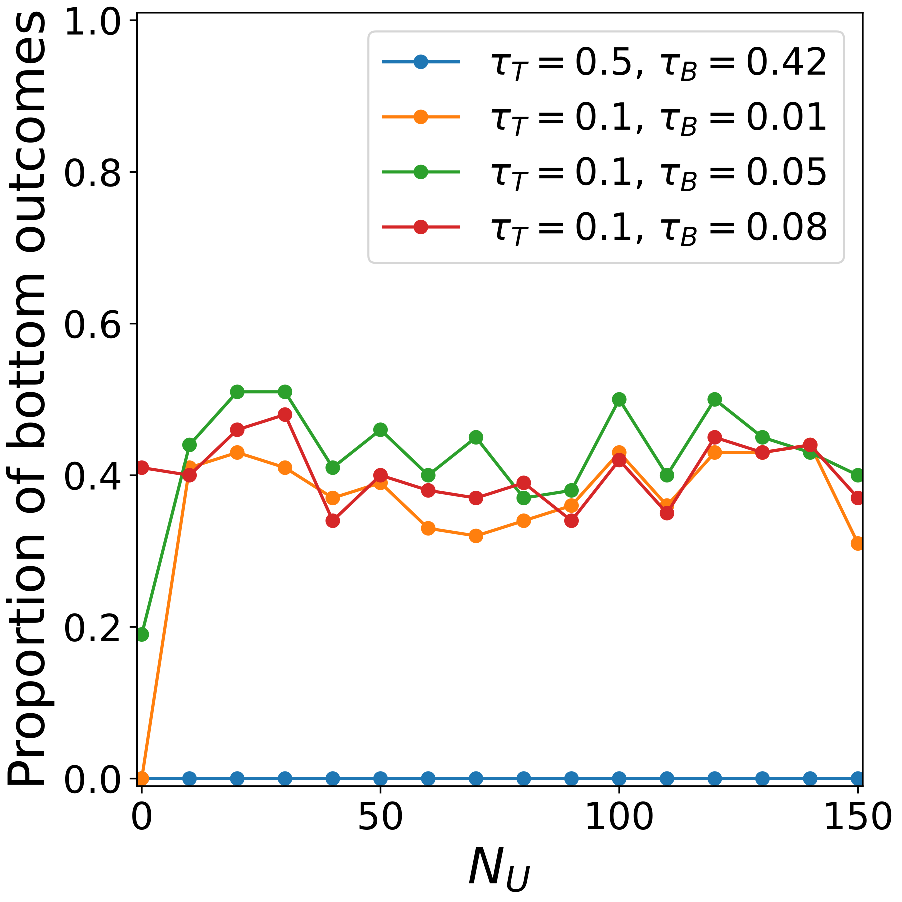}
        \caption{}
        \label{fig: percentages-b}
    \end{subfigure}
    \caption{(a) Frequency of each possible outcome of Case 3 described in Figure \ref{fig:three-pop-4} averaged over 200 runs. (b) Proportion of bottom outcomes for different number of uninformed individuals when $N_L=20$ and $N_F=50$. Each point corresponds to the average over 100 runs with different initial conditions.}\label{fig: percentages}
\end{figure}

To quantify which of these three outcomes is more dominant we count the frequency of each outcome out of $200$ runs with different initial conditions chosen uniformly at random. Let $\bar{\bv}=(\bar{v}_x,\bar{v}_y)$ be the mean velocity at the final time $T$. We classify bottom outcomes as those satisfying $\bar{v}_x,\bar{v}_y<-0.75$, top outcomes as $\bar{v}_x,\bar{v}_y>0.75$, and  mill outcomes are classified using the momentum defined in \eqref{eq: pol and mom} at the final time, and have to satisfy $M(T)>0.75$ (see Figure \ref{fig:three-pop-4c-quantities}). Figure \ref{fig: percentages-a} summarizes the outcome frequencies. Approximately $43\%$ of the simulations result in the flock moving toward the bottom target, while top-directed flocks and mills each occur in about $20\%$ of cases.
Directed flocks moving in other directions (e.g. parallel to the $y$-axis, or in any other direction outside these ranges) are included in Figure \ref{fig: percentages-a} as ``Other''. Of these 37 "Other" cases, $43\%$ move toward the first quadrant, $24\%$ to the second, $11\%$ to the third, and $22\%$ to the fourth.

Finally, we examine the effect of increasing the number of uninformed individuals in Figure \ref{fig: percentages-b}, across different values of $\tau_T$ and $\tau_B$. When the leaders' conviction is  too strong ($\tau_T = 0.5$, blue line), the bottom target is never selected, regardless of how many uninformed individuals are included. In contrast, when the leaders' conviction is mild, the influence of the uninformed becomes apparent—particularly for $N_U = 10,\ 20,\ 30,\ 40$, where the proportion of bottom target selections is around $0.4$. This behaviour is consistent with the trends observed in Figure \ref{fig: percentages-a}. Beyond approximately $N_U = 50$, the effect of the uninformed appears to stabilise and remains nearly constant.

\section{Conclusions}

In this paper we proposed different individual-based models for decision making in animal groups. Our results provide further insights into leadership and decision making in heterogeneous groups of leaders, followers and uninformed individuals. The proposed models combine swarming dynamics with continuous in time opinion evolution - they are capable of reproducing biological observations, particularly those by Couzin et al. \cite{couzin2002collective}, but also provides analytical insights into the mechanisms driving collective decision-making in spatially distributed systems.

One of the central contributions corresponds to quantifying the effect of uninformed individuals in heterogenous groups. We show, using both analytical insights as well as extensive numerical simulations, that these agents, despite lacking directional or opinion bias, influence the decision making in groups significantly. Their neutral stance effectively mitigates the dominance of strongly opinionated leaders, allowing the preference of the numerical majority (followers) to prevail. This was evident even when uninformed individuals had no active role in updating or transmitting opinions, underscoring their passive yet stabilising role.

The three-population model developed in this paper, distinguishing leaders, followers, and uninformed individuals, adds a new layer of realism to classical opinion models by incorporating spatial interaction, self-propulsion, and bounded confidence mechanisms for opinions. Analytical results for simplified versions of the model match numerical simulations, validating the consistency of the framework. Furthermore, the identification of steady-state behaviours, such as flocking, depending on population parameters and interaction rules, highlights the model's ability to capture complex emergent phenomena.

The findings have broader implications for the design of decentralised decision-making systems in robotics, social networks, and collective animal behaviour. In particular, the inclusion of uninformed agents emerges as a potentially simple yet powerful mechanism to enhance democratic outcomes and prevent leader-driven polarisation.

Future work could explore time-dependent conviction parameters, more complex network topologies, and stochastic influences, further aligning the model with realistic scenarios in socio-biological and artificial systems. Another important direction is the derivation of mean-field (continuum) models that capture the macroscopic behaviour of large populations. Such models would not only provide a rigorous connection between microscopic interactions and emergent group-level dynamics but also facilitate analytical treatment and reduce computational complexity for large-scale simulations.

\bibliographystyle{abbrv}
\bibliography{references.bib}

\section{General swarming model}\label{app: swarming model}

This section gives a short overview of the observed patterns of D'Orsogna et al, see \cite{d2006self} as well as the coupled opinion-swarming model for different parameter regimes.\\

\begin{figure}[ht!]
    \centering
    \includegraphics[width=\linewidth]{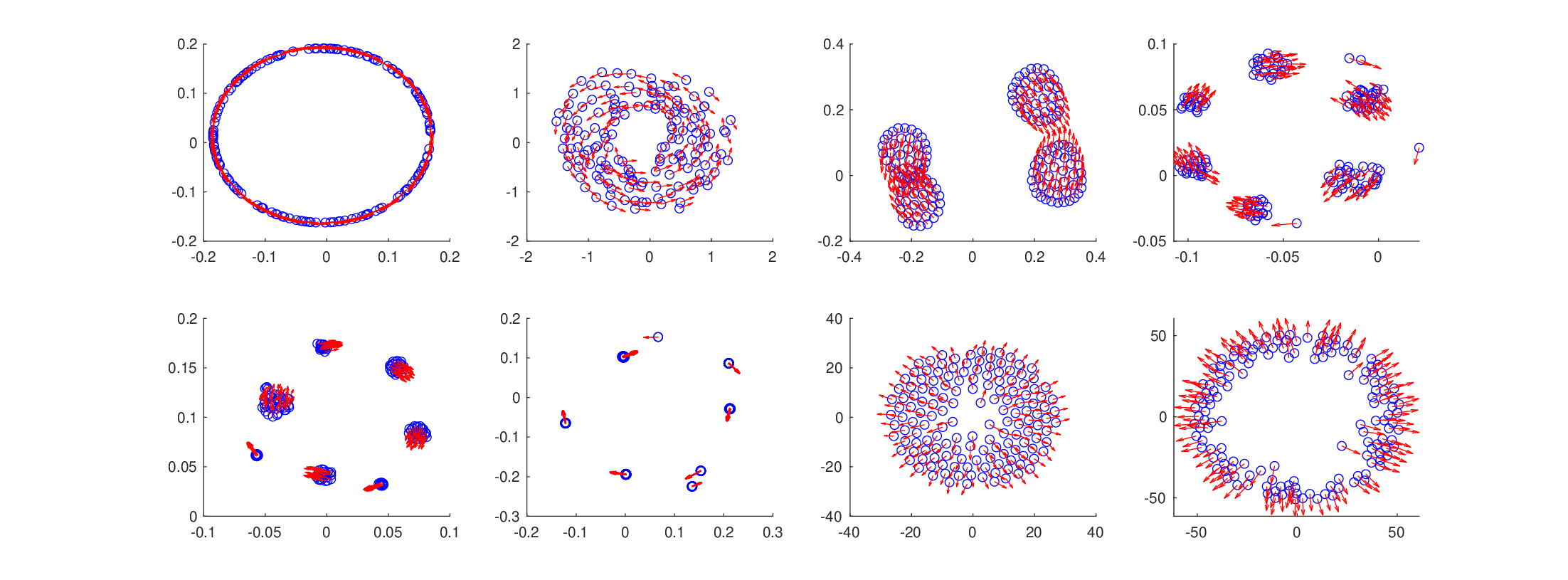}
    \includegraphics[width=\linewidth]{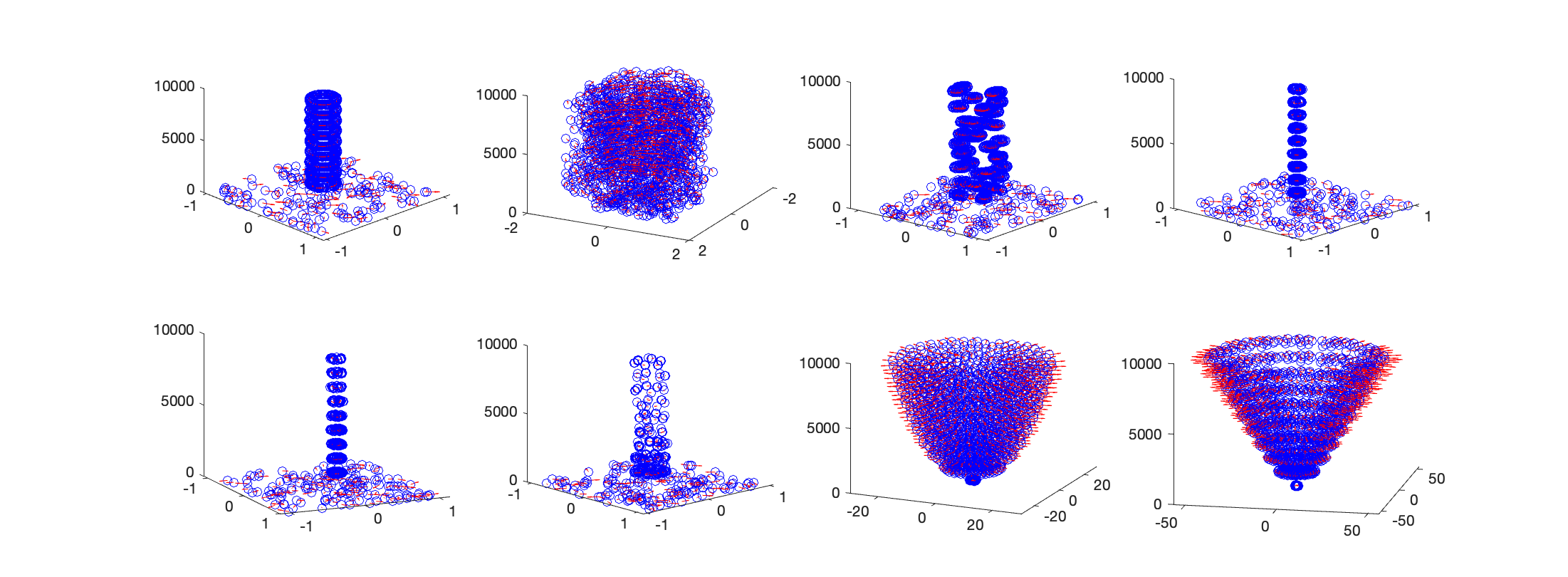}
    \caption{Classical results without opinion}
    \label{fig: class swarming}
\end{figure}

\begin{figure}[ht!]
    \centering
    \includegraphics[width=\linewidth]{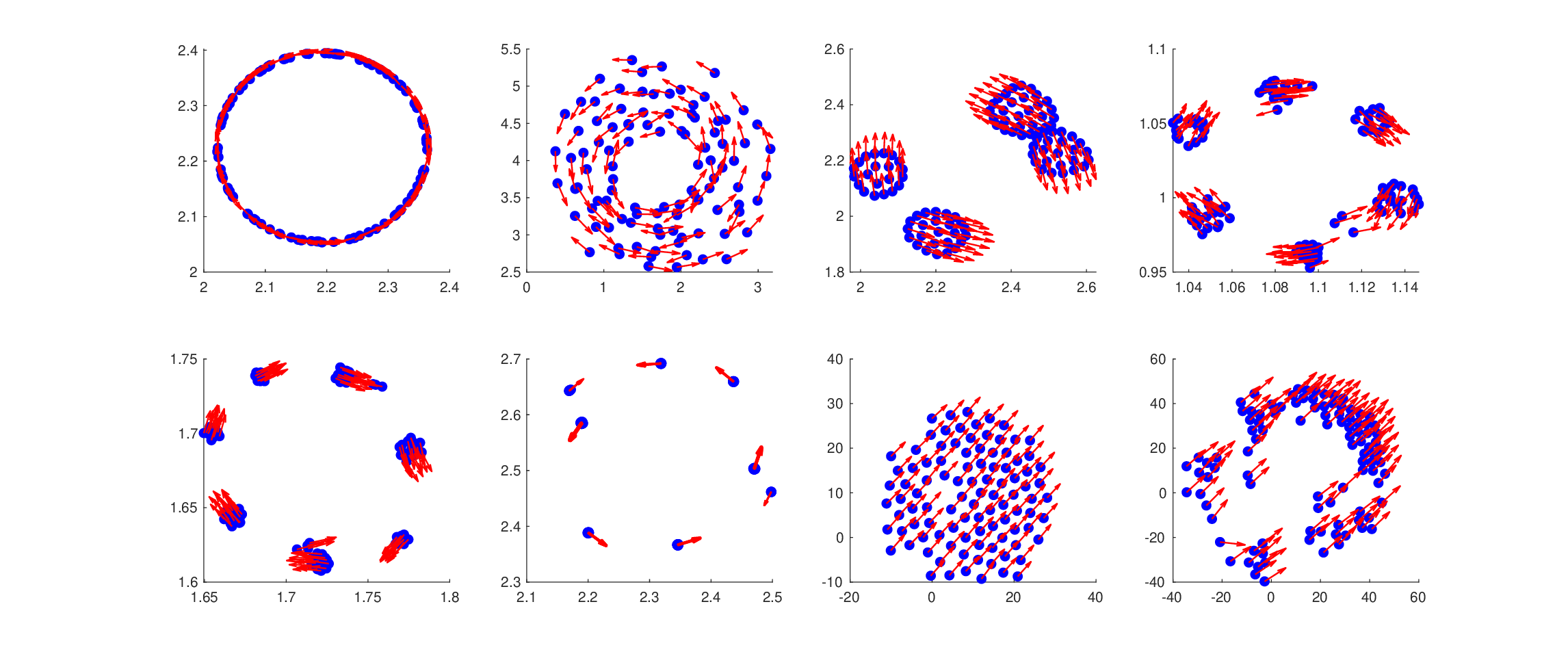}
    \includegraphics[width=\linewidth]{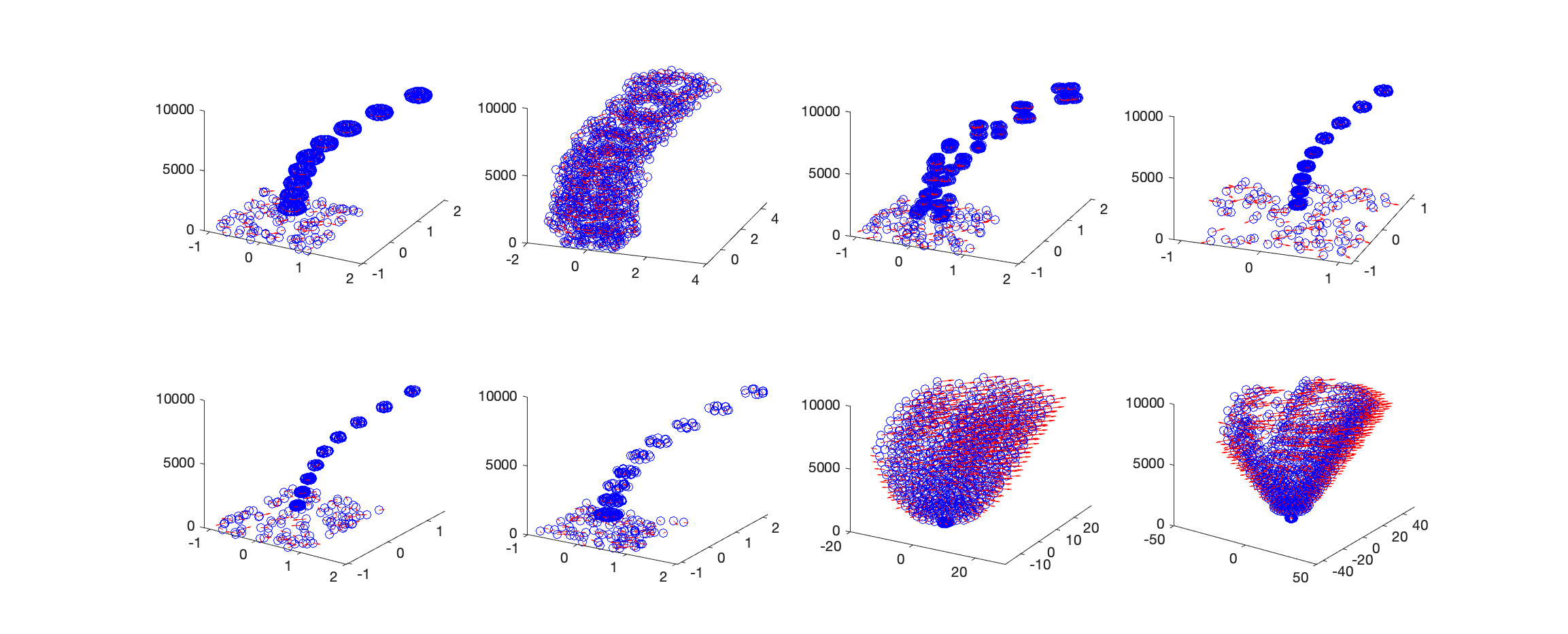}
    \caption{Swarming with opinions}
    \label{fig: swarming with op 1pop}
\end{figure}

We recall the patterns reported in \cite{d2006self} in Figure \ref{fig: class swarming} - such as rotating mills or flocks - for the parameter regimes reported in Table \ref{tab: parameters flocking}. Figure \ref{fig: swarming with op 1pop} shows the respective patterns of the proposed opinion swarming model \eqref{eq: one-population} (using the parameters listed in Table \ref{tab: parameters flocking}) for the case $\tau_B = 0$ and $\tau_T = 1$. We recall that individuals only have a preference towards the top target and tend to align their velocity with $\mathbf{v}_T = (1,1)$. The coefficient controlling the velocity alignment due to opinion, $\gamma_s$ is chosen as $\gamma_T=\gamma_B=1$. \\
The first two rows of Figure \ref{fig: swarming with op 1pop} display the final velocity configurations, which appear very similar to those obtained without opinions. However, when examining the time evolution (last two rows), we observe that the entire system, while maintaining the same configuration, gradually drifts toward the positive velocity direction, driven by the preferred opinion $w_T = 1$.\\
To show the influence of the opinion on the different configurations in Figure \ref{fig: swarming with op 1pop}, we are going to study more carefully the dynamics corresponding to the parameters II and VIII in Table \ref{tab: parameters flocking}. The results presented here are for $\gamma_T=\gamma_B=1$, $r_w=0.5,\ 1$, $r_\bx=1$, $\tau_T=0.05$ and $\tau_B=0$. 

\begin{table}[ht]
    \centering
    \begin{tabular}{|c|c|c|c|c|c|c|}
        \hline
           & $\alpha$ & $\beta$ & $C_a$  & $\ell_a$ & $C_r$ & $\ell_r$ \\
        \hline\hline
          I & $1$ & $0.5$ & $100$ & $1$ & $50$ & $0.5$\\
        \hline
          II & $1$ & $0.5$ & $50$ & $1$ & $60$ & $0.5$\\
        \hline
         III & $1$ & $0.5$ & $100$ & $1$ & $60$ & $0.5$ \\
        \hline
          IV & $1$ & $0.5$ & $100$ & $1$ & $40$ & $6$\\
        \hline
          V & $1$ & $0.5$ & $100$ & $1$ & $50$ & $1.2$\\
        \hline
          VI & $1$ & $0.5$ & $100$ & $1$ & $60$ & $0.7$\\
        \hline
         VII  & $0.1$ & $5$ & $100$ & $1.2$ & $350$ & $0.8$\\
        \hline
         VIII  & $1$ & $5$ & $100$ & $1.2$ & $350$ & $0.8$\\
        \hline
    \end{tabular}
    \caption{Example of a table with an extra empty left column}\label{tab: parameters flocking}
\end{table}

\end{document}